\documentclass[
reprint, prb,
superscriptaddress,
 amsmath,amssymb,
 aps,amsthm
]{revtex4-2}

\usepackage{graphicx}
\usepackage{dcolumn}
\usepackage{bm}
\usepackage{physics}
\usepackage{tikz}
\usepackage{algorithm}
\usepackage{algpseudocode}
\usepackage{subcaption}
\usepackage{hyperref, soul}
\hypersetup{linktocpage,,colorlinks=true}
\usepackage{physics}

\newtheorem{remark}{Remark}

\begin{document}

\preprint{APS/123-QED}

\title{
Magic Entropy in Hybrid Spin-Boson Systems
}

 

\author{Samuel Crew}
\thanks{SC, YLL and HHL contributed equally to the work}
\email{samuel.c.crew@gmail.com}
\affiliation{Department of Physics, National Tsing Hua University, Hsinchu 30013, Taiwan}

\author{Ying-Lin Li}
\email{s1012424@gmail.com}
\affiliation{Department of Physics, National Tsing Hua University, Hsinchu 30013, Taiwan}

\author{Heng-Hsi Li}
\email{steven0823255219@gmail.com}
\affiliation{Department of Physics, National Tsing Hua University, Hsinchu 30013, Taiwan}

\author{Po-Yao Chang}
\email{pychang@phys.nthu.edu.tw}
\affiliation{Department of Physics, National Tsing Hua University, Hsinchu 30013, Taiwan}
\affiliation{Yukawa Institute for Theoretical Physics, Kyoto University, Kyoto 606-8502, Japan}

\date{\today}

\begin{abstract}
We introduce entropic measures to quantify non-classical resource in hybrid spin-boson systems. We discuss the stabilizer R\'enyi entropy in the framework of phase space quantisation and define an analogous hybrid magic entropy and a mutual magic entropy that capture the distribution of quantum magic across spin and bosonic subsystems. 
We use these entropic measures to demonstrate two key phenomena: the detection of the superradiant phase transition in the Dicke model and the dynamics of magic in the Jaynes-Cummings model following a quench. We develop a Monte Carlo numerical scheme to enable practical computation in many-body examples.
\end{abstract}

\maketitle

\section{Introduction}
Entanglement has long been recognised as a key feature distinguishing quantum and classical correlations. However, it is not the only resource that captures quantum complexity. Another crucial notion is non-stabilizerness, or magic \cite{bravyi2005universal}, which quantifies the extent to which a qubit quantum state or protocol deviates from classical simulability; as characterised by the Gottesman-Knill theorem \cite{gottesman1997stabilizer}. Magic states may be viewed as a resource that enables universal quantum computation. Quantum resource theories~\cite{Chitambar_2019} provide a general framework for classifying operations and states as either free or resourceful. For example, in the theory of entanglement, the free operations are local operations and classical communication, while in magic resource theories, the free operations are Clifford operations and measurements.  

In recent years, several measures of non-stabilizerness have been proposed~\cite{veitch2014resource,raussendorf2017contextuality,veitch2012negative,mari2012positive,garcia2017geometry,beverland2020lower}. However, evaluating many of these measures for many body ground states is computationally challenging. A particularly notable recent proposal is the \emph{stabilizer R\'enyi entropy}~\cite{leone2022stabilizer}, a magic monotone \cite{leone2024stabilizer} that has attracted interest for its computability (enjoying both efficient classical and quantum algorithms \cite{stratton2025algorithm,haug2024efficient,liu2025nonequilibrium,lami2023nonstabilizerness,ding2025evaluating,tarabunga2023many}) and since its operational interpretation enables experimental measurement~\cite{oliviero2022magic}.

In this work, we extend the stabilizer R\'enyi entropy to hybrid systems composed of both discrete (spin) and continuous-variable (bosonic) degrees of freedom. The resulting entropic measure captures quantum resources arising from both non-stabilizerness and non-Gaussianity. We begin by revisiting the stabilizer R\'enyi entropy from the perspective of the geometric quantisation of phase space, drawing parallels between Clifford and Gaussian resource theories. This framework naturally leads to a continuous variable analogue, \emph{Gaussian R\'enyi entropy}, for bosonic systems and a spin-boson measure, \emph{hybrid magic R\'enyi entropy} for hybrid systems. Finally, we introduce a \emph{mutual magic entropy} that quantifies how these two forms of quantum `magic' are correlated across subsystems.

Spin-boson models describe spins interacting with bosonic modes, capturing the physics of a wide range of systems. In particular, the Dicke model \cite{dicke1954coherence} and its variants model spins coupled to a common cavity mode and have been experimentally realised in, for example, quantum optics~\cite{Baumann2010}, superconducting qubits~\cite{Niemczyk2010, Karyn2016}, and trapped ions~\cite{Porras2004, Porras2008}. These experimental platforms enable quantum simulations of both equilibrium and dynamical phenomena, including the superradiant phase transition. While the entanglement structure of the Dicke model is well-understood~\cite{lambert2004entanglement,lambert2005entanglement}, the nature and dynamics of magic shared between spin and bosonic degrees of freedom remains largely unexplored. 

To shed new light on this questions, we apply our hybrid magic resource entropy to the ground states of the Dicke model and demonstrate that they detect many-body phase transitions. We find that the superradiant phase transition manifests as a maximum in both the non-Gaussian and non-stabilizer entropies. Moreover, the mutual entropy also reaches a maximum, indicating that quantum resources are maximally correlated between the spin and bosonic sectors. We then study the dynamics of magic in the Jaynes–Cummings model \cite{jaynes2005comparison} (a variant of the Dicke model with one spin) after a quench and observe that the stabilizer R\'enyi entropy has distinct dynamics compared with its Gaussian and hybrid analogues.

In practice, to efficiently compute hybrid magic entropies of many-body models, we develop a Monte Carlo method and benchmark its performance against exact diagonalisation. In addition, we derive analytical expressions for the entropies using perturbation theory in the weak-coupling regime, finding good agreement between the numerical and analytical results.

\paragraph*{Outline.}
The paper is organized as follows. In Section~\ref{sec:stabtheory}, we review stabilizer theory in both discrete and continuous-variable settings. Section~\ref{sec:resourceentropy} introduces the resource entropies from a geometric and phase-space perspective. In Section~\ref{sec:Examples}, we apply these ideas to the Dicke model, presenting both perturbative and numerical results. We conclude in Section~\ref{sec:discussion} with a discussion of future directions.

\section{Stabilizer theory}\label{sec:stabtheory}
We begin with a review of the basic elements of stabilizer theory as applied to both discrete qubit systems and continuous-variable (CV) systems. In discrete systems, stabilizer states arise as the orbit of the Clifford group acting on the computational basis zero state, and their structure is governed by the Pauli group and its normaliser. In the CV setting, the analogous role is played by Gaussian states and symplectic transformations acting on phase space. In this section we set the necessary background and notation, while also emphasising analogies between the discrete and continuous settings that motivate our phase space discussion in section \ref{sec:resourceentropy} that follows.

\subsection{Discrete qubit systems}
Let $\mathcal{H} = (\mathbb{C}^2)^{\otimes N}$ denote the Hilbert space of a system of $N$ qubits. We label computational basis states by $|x\rangle$, where $x \in \mathbb{F}_2^N$. We begin by introducing the Pauli group, defined as the group generated by tensor products of the single-qubit Pauli matrices $X$, $Y$, and $Z$, acting on the $N$ qubits. It is given by  
\begin{equation}
\begin{split}
\mathcal{P}_N &:= \langle \{ I, X, Y, Z \}^{\otimes N} \rangle \\
&= \left\{ i^k\, \sigma_1 \otimes \cdots \otimes \sigma_N \;\middle|\; k \in \mathbb{Z}_4,\ \sigma_i \in \{ I, X, Y, Z \} \right\}.
\end{split}
\end{equation}
The Clifford group is defined as the normaliser (up to phases) of the Pauli group inside the group of unitary matrices
\begin{equation}
    \text{Cliff}_N := \{ U \in U(2^N) \, : \, U \mathcal{P}_N U^{\dag} \subset \mathcal{P}_N \},
\end{equation}
It is generated by the Hadamard gate, phase gate and CNOT gate.
Stabilizer states are defined as the orbit of $|0 \rangle$ under the Clifford group $\text{Cliff}_N$:
\begin{equation}
    \text{Stab}_N := \{ U |0\rangle^{\otimes N} \, | \, U \in \text{Cliff}_N\}.
\end{equation}
The Gottesman-Knill theorem \cite{gottesman1997stabilizer,gottesman1998heisenberg,aaronson2004improved} implies that, despite the potentially significant entanglement introduced by the CNOT gate, the dynamics of stabilizer states and Pauli measurements are not only sub-universal but also essentially classical; they may be efficiently simulated on a classical computer.

\subsection{Continuous variable systems}
We now consider a system of $N$ bosonic degrees of freedom, described by the Hilbert space $\mathcal{H}_N = L^2(\mathbb{R}^N)$. The space admits a Fock space description given by the symmetric tensor product of the single-mode Hilbert spaces as:
\begin{equation}
\mathcal{H}_N = \bigoplus_{n_1, \dots, n_N \geq 0} \mathrm{Sym}^{n_1}(L^2(\mathbb{R})) \otimes \dots \otimes \mathrm{Sym}^{n_N}(L^2(\mathbb{R})).
\end{equation}
The Hilbert space has a basis of number eigenstates
\begin{equation}
|n_1, \dots, n_N\rangle = \frac{1}{\sqrt{n_1! \dots n_N!}} \left( \hat{a}_1^\dagger \right)^{n_1} \dots \left( \hat{a}_N^\dagger \right)^{n_N} |0\rangle,
\end{equation}
where $\{\hat{a}_i\}_{i=1}^N$ are creation and annihilation operators satisfying 
\begin{equation}
[\hat{a}_i, \hat{a}_j^\dagger] = \delta_{ij}, \quad
[\hat{a}_i, \hat{a}_j] = 0, \quad
[\hat{a}_i^\dagger, \hat{a}_j^\dagger] = 0.
\end{equation}
and $|0\rangle$ is the Fock vacuum annihilated by all $\hat{a}_i$ for $i=1,\ldots,N$. It is convenient to also define the quadrature operators $\hat{x} = (\hat{q}_1,\hat{p}_1,\ldots \hat{q}_N, \hat{p}_N)^T$ with $\hat{q}_j = (\hat{a}_j + \hat{a}_j^{\dag})/\sqrt{2}$ and $\hat{p}_j = i(\hat{a}_{j} - \hat{a}_{j}^{\dag})/\sqrt{2}$.

The analogue of the Pauli group is the Heisenberg-Weyl group, denoted $\text{HW}_N$, which is isomorphic to the group generated by the displacement operators 
\begin{equation}\label{eq:CVdisplacement}
    \hat{D}(\xi) := \exp\left( -i \xi^T \Omega \hat{x} \right),
\end{equation}
where $\xi \in \mathbb{R}^{2N}$ and $\Omega$ denotes the standard sympletic form on $\mathbb{R}^{2N}$. The analogue of the qubit Clifford group is defined as the normaliser of $\text{HW}_N$ inside $U(L^2(\mathbb{R}^N))$ and is concretely realised by the Gaussian unitary operators. These are most conveniently described in the Heisenberg picture where Gaussian unitaries act on the quadrature operators by
\begin{equation}
   U_{S,d}:\hat{x} \to S \hat{x} + d,
\end{equation}
with $S$ a $2N \times 2N$ symplectic matrix and $d$ a $N$-dimensional displacement vector. 
We denote this group of Gaussian unitaries by
\begin{equation}
    G_N = \{U_{S,d} \,: \, S \in \text{Sp}(2n, \mathbb{R}), \,d \in \mathbb{R}^N\},
\end{equation}    
examples of Gaussian unitaries include the displacement operators, squeezing operators, phase shift and beam splitters. Finally, Gaussian states are similarly constructed from the action of $G_N$ on the Fock vacuum $|0 \rangle$.
The bosonic analogue of the Gottesman-Knill theorem \cite{bartlett2002efficient} states that Gaussian operations and measurements are similarly classically efficient to simulate. For a more thorough introduction to CV systems we refer the reader to the reviews \cite{walschaers2021non,weedbrook2012gaussian}.

\section{Magic resource entropy}\label{sec:resourceentropy}
As discussed in the previous section, non-Clifford resources are essential for achieving quantum computational advantage~\cite{campbell2017roads}. The amount of such non-Clifford resource is colloquially referred to as \emph{magic}~\cite{bravyi2005universal}. The resource theory of magic~\cite{howard2017application, howard2014contextuality,Chitambar_2019} provides a formal framework for quantifying non-Cliffordness in quantum states or protocols. Within this theory, Clifford operations are deemed free, while the preparation of magic states represents a costly resource. 

A \emph{magic monotone} is any real-valued function on quantum states that is non-increasing under stabilizer operations, thereby providing a meaningful quantifier of the non-Clifford content of a state. These monotones play a role analogous to entanglement monotones in the resource theory of entanglement. Several magic monotones have been introduced in the literature, examples include the mana \cite{veitch2014resource}, Wigner function negativity measures \cite{raussendorf2017contextuality, veitch2012negative, mari2012positive}, stabilizer rank \cite{garcia2017geometry}, and stabilizer nullity~\cite{beverland2020lower}. 

In this work, however, we focus on a particular monotone: the \emph{stabilizer R\'enyi entropy}, recently introduced by Leone \textit{et. al.} \cite{leone2022stabilizer}. A key advantage of this measure is that it admits a direct operational interpretation, avoiding the need for optimisation procedures typical of other magic monotones.

We begin by reviewing the necessary aspects of geometric quantisation (see~\cite{de2006symplectic} for a significantly more comprehensive account) and formulate the stabilizer Rényi entropy in this broader setting. We then revisit the structure of stabilizer and Gaussian states in qubit and continuous-variable (CV) systems, highlighting their formal analogies. In subsection~\ref{subsec:spinbosonmagic}, we extend the entropy to hybrid spin-boson systems and present a practical numerical scheme for its evaluation. Finally, we introduce a mutual resource measure that quantifies how magic resource is correlated across the spin and bosonic subsystems.

\subsection{Phase space and geometric quantisation}\label{subsec:geometricquantisation}
Let us consider $(\mathcal{M},\omega)$ a locally compact phase space where $\mathcal{M}$ has an additive Abelian group structure, for example a vector space over a (possibly finite) field, and $\omega$ is a symplectic form. We denote phase space points by $\xi \in \mathcal{M}$. Local compactness ensures the existence of a Haar measure $\mathrm{d}{\mu}(\zeta)$. As an additive group, the phase space $\mathcal{M}$ acts on itself by translation with generators denoted $D(\zeta): \eta \to \eta + \zeta$.

We consider geometric quantisations of $\mathcal{M}$ that provide a Hilbert space $\mathcal{H}$ carrying a \textit{projective} unitary representation of the translation group. That is displacement (or Weyl) operators $\hat{D}(\zeta): \mathcal{H} \to \mathcal{H}$ satisfying
\begin{equation}\label{eq:projectiverep}
    \hat{D}(\zeta)\hat{D}(\zeta') = e^{i\omega(\zeta,\zeta')}\hat{D}(\zeta + \zeta').
\end{equation}
We suppose further that the operators $\hat{D}(\zeta)$ form an (over) complete set for operators on $\mathcal{H}$, normalised so that $\hat{D}(0)=\mathbb{I}$. Hence a trace-class operator $\hat{A}$ may be expressed as \footnote{This equation fixes the normalisation of the Haar measure on $\mathcal{M}$.}
\begin{equation}\label{eq:resolnidentity}
    \hat{A} = \int_{\mathcal{M}} \mathrm{d}\mu(\zeta) \, \tr(\hat{A} \hat{D}(\zeta)) \hat{D}^{\dag}(\zeta).
\end{equation}
In the case $A=I$ we see that
\begin{equation}\label{eq:tracenormalisation}
    \tr \hat{D}(\zeta) = \delta_{\mathcal{M}}(\zeta),
\end{equation}
where $\delta_{\mathcal{M}}(\zeta)$ is the normalised (relative to the Haar measure) distribution satisfying $1=\int_{\mathcal{M}}\mathrm{d}\mu(\zeta) \, \delta_{\mathcal{M}}(\zeta)$.

\subsubsection{Weyl function entropy}
Let us consider a quantum state $\rho$ on $\mathcal{H}$. The operator $\rho$ is trace class and so the resolution of the identity in equation \eqref{eq:resolnidentity} applies and we have
\begin{equation}
    \rho = \int_{\mathcal{M}} \mathrm{d}\mu(\zeta) \,\chi_{\rho}(\zeta) \hat{D}^{\dag}(\zeta),
\end{equation}
where $\chi_{\rho}: \mathcal{M} \to \mathbb{C}$ denotes the Weyl function of a state defined by
\begin{equation}
    \chi_{\rho}(\zeta) := \tr \rho \hat{D}(\zeta).
\end{equation}
Now let us consider the purity $\tr \rho^2$. The group structure identity \eqref{eq:projectiverep} together with \eqref{eq:tracenormalisation} implies that
\begin{equation}
    \tr \hat{D}(\zeta)\hat{D}(\zeta') = \delta_{\mathcal{M}}(\zeta-\zeta'),
\end{equation}
and hence $\tr \rho^2 = \int \mathrm{d}\mu(\zeta) \, |\chi_{\rho}(\zeta)|^2$. The result is that we may define a probability distribution $p_{\rho}: \mathcal{M} \to \mathbb{R}$ by
\begin{equation}\label{eq:generalprobability}
    p_{\rho}(\zeta) := \frac{|\chi_{\rho}(\zeta)|^2}{\tr \rho^2}\mu.
\end{equation}
When the state is pure, $\rho = |\psi \rangle \langle \psi|$, this probability distribution is proportional to the square Weyl function. We may then consider an associated $\alpha$-R\'enyi entropy of the distribution $p_{\rho}$ and define
\begin{equation}\label{eq:generalentropy}
    H_{\alpha}(\rho) := \frac{1}{1-\alpha} \log \int_{\mathcal{M}} \mathrm{d}\zeta \,p_{\rho}(\zeta)^{\alpha},
\end{equation}
for an arbitrary state $\rho$.

\subsubsection{Entropy properties}\label{subsubsec:entropyproperties}
We now discuss some properties of the resource entropy of a state $\rho$ defined in \eqref{eq:generalentropy}. In this section we focus on the case of pure states $\rho = |\psi \rangle \langle \psi|$. We first consider invariance under stabilizer operations. Suppose that $S$ is an automorphism of $\mathcal{M}$ satisfying
\begin{equation}
    S D(\zeta) S^{-1} = D(S \cdot \zeta),
\end{equation}
under quantisation we have that
\begin{equation}
    \hat{S} \hat{D}(\zeta) \hat{S}^{-1} = e^{i \phi} \hat{D}(S\cdot \zeta),
\end{equation}
so that $S$ normalises the displacement group up to a phase. We have $\mathrm{d}\mu(\zeta) = \mathrm{d}\mu(S\cdot \zeta)$ hence changing variables in the integral \eqref{eq:generalentropy} we see that the entropy is preserved under stabilizer operations \footnote{By definition, those that normalize the displacement group up to phase.}.
The property of additivity under tensor products $H_{\alpha}(|\psi_1\rangle \otimes |\psi_2 \rangle) = H_{\alpha}(|\psi_1\rangle) + H_{\alpha}(|\psi_2 \rangle)$ is also immediate.

\subsection{Stabilizer R\'enyi entropy}
Let us now specialise to the case $\mathcal{M} =  \mathbb{F}_2^{2N}$, the classical phase space of $N$ qubits. The quantised displacement operators $\hat{D}(a,b)$ are realised by Pauli strings acting on $\mathcal{H} = (\mathbb{C}^2)^{\otimes N}$ and given explicitly by
\begin{equation}
    \hat{D}(a,b) = \sigma_{a_1,b_1} \otimes \ldots \otimes \sigma_{a_N,b_N}
\end{equation}
where $\zeta = (a,b) = (a_1,\ldots,a_N,b_1,\ldots b_N) \in \mathcal{M}$ and $\sigma_{a,b} := i^{ab}X^aZ^b$. The construction of section \ref{subsec:geometricquantisation} yields a probability distribution on phase space $p_{\rho}: \mathbb{F}_2^{2N} \to \mathbb{C}$ associated to a state $\rho$. With the normalisation factor in equation \eqref{eq:generalprobability} given by $\mu = 1/2^N$.

The stabilizer R\'enyi entropy of a state $\rho$, denoted $M^{S}(\rho)$, is defined to be the $\alpha$-R\'enyi entropy \eqref{eq:generalentropy} shifted so that $M_{\alpha}^S(\rho) = 0$ if and only if $\rho$ is a stabilizer state. Explicitly,
\begin{equation}\label{eq:SRE}
    M_{\alpha}^S(\rho) := H_{\alpha}(p_{\rho}) + (1-\alpha)S_2(\rho) - N \log 2,
\end{equation}
where $H$ is the entropy function \eqref{eq:generalentropy} and $S_2(\rho)$ denotes the $2$-R\'enyi entropy of $\rho$. The integral here is over the counting measure on phase space \textit{i.e.} in this case it is a finite sum over Pauli strings.

The invariance of the entropy under stabilizer operations follows from the fact, reviewed in \textit{e.g.} lemma 2.1 of \cite{gross2021schur}, that for every Clifford unitary $U \in \text{Cliff}_N$ there exists a $\Gamma \in \text{Sp}(2N,\mathbb{F}_2)$ such that
\begin{equation}
    U \sigma_{a,b} U^{\dag} = \sigma_{\Gamma \cdot (a,b)},
\end{equation}
and we may apply the general invariance argument of the previous section \ref{subsubsec:entropyproperties}.

The stabilizer R\'enyi entropies with $\alpha \ge 2$ are in addition magic monotones for pure states $\rho = |\psi \rangle \langle \psi |$, the proof may be found in \cite{leone2024stabilizer}, and there exists a number of quantum \cite{stratton2025algorithm,haug2024efficient} and classical \cite{liu2025nonequilibrium,lami2023nonstabilizerness,ding2025evaluating,tarabunga2023many} algorithms to efficiently compute the stabilizer R\'enyi entropy.

\subsection{Gaussian R\'enyi entropy}
We now turn to the case of a continuous phase space $\mathcal{M} = T^*\mathbb{R}^N \cong \mathbb{R}^{2N}$. The quantisation we consider is then a CV system of $N$ bosons with Hilbert space $\mathcal{H} = L^2(\mathbb{R}^N)$. We previously defined the displacement operators $\hat{D}(\xi)$, with $\xi \in \mathcal{M}$, in equation \eqref{eq:CVdisplacement} and note here only in addition that they satisfy the projective representation condition \eqref{eq:projectiverep}.

The construction of section \ref{subsec:geometricquantisation} then yields a probability density on phase space $p: \mathbb{R}^{2N} \to [0,1]$ with normalisation factor given by $\mu = 1/\pi^N$. We may now define the Gaussian R\'enyi entropy of a state in terms of the  $\alpha$-R\'enyi entropy of this probability distribution by
\begin{equation}
\label{eq:GRE}
    M_{\alpha}^G(\rho) := H_{\alpha}(p_{\rho}) - (1-\alpha) S_2(\rho) - N \log \pi + \frac{N}{1-\alpha}\log \alpha,
\end{equation}
In the limit $\alpha \to 1$ we recover the differential entropy and write $M^G(\rho)$ with no subscript for this case. The linear shift of the Gaussian R\'enyi entropy is chosen such that $M_{\alpha}^G(\rho) = 0$ when $\rho$ is a Gaussian state.

\begin{remark}\label{remark:measurement}
We conclude the section by noting that, analogous to the Bell measurement interpretation~\cite{montanaro2017learning} of the stabilizer Rényi entropy in qubit systems, the Gaussian R\'enyi entropy can, formally, be expressed in terms of measurement statistics. We define the following projectors parametrised by $\xi \in \mathcal{M}$:
\begin{equation}
P_{\xi} := (\hat{D}(\xi) \otimes I)\, |\Phi\rangle \langle \Phi|\, (\hat{D}(\xi) \otimes I),
\end{equation}
where $|\Phi\rangle$ denotes the (unnormalisable) maximally entangled state $|\Phi\rangle = \int_{\mathbb{R}^N} \mathrm{d}x\, |x\rangle \otimes |x\rangle$
which may be approximated by a two-mode squeezed vacuum state.

Now, in terms of the replica state $|\varphi\rangle := |\psi\rangle \otimes |\psi^*\rangle \in \mathcal{H}^{\otimes 2}$, the probability distribution \eqref{eq:generalprobability} may be written as the probability of obtaining outcome $\xi$ in a projective measurement, namely
\begin{equation}
|\chi_\psi(\xi)|^2 = \tr(P_\xi\, |\varphi\rangle \langle \varphi|).
\end{equation}
The Gaussian Rényi entropy~\eqref{eq:GRE} is then the Rényi entropy of this measurement probability distribution.
\end{remark}

\begin{remark}
    Continuous variable generalisations of the stabilizer R\'enyi entropy also appear in recent work \cite{hahn2025bridging} where quantitative relations between magic and non-Gaussianity are discussed.
\end{remark}

\subsection{Hybrid magic R\'enyi entropyy}\label{subsec:spinbosonmagic}
We now consider a system with $M$ bosonic and $N$ spin degrees of freedom and a classical phase space of the product form $\mathcal{M} = \mathbb{R}^{2M} \times \mathbb{F}_2^{2N}$. The quantisation of this system gives rise to a Hilbert space  $\mathcal{H} = \mathcal{H}_b \otimes \mathcal{H}_s$ with $\mathcal{H}_b = L^2(\mathbb{R}^M)$ and $\mathcal{H}_s = (\mathbb{C}^2)^{\otimes N}$. We consider a general pure (but not necessarily separable) state $|\psi \rangle \in \mathcal{H}$ with an associated Weyl function $\chi_{\psi}: \mathbb{R}^{2M} \times \mathbb{F}_2^{2N} \to \mathbb{C}$ given explicitly by
\begin{equation}
    \chi_{\psi}(\zeta) := \langle \psi | \sigma_{a,b} \otimes \hat{D}(\xi) |\psi \rangle,
\end{equation}
where the phase space coordinate is denoted $\zeta = (a,b,\xi)$ with $(a,b) \in \mathbb{F}_2^{2N}$ and $\xi \in \mathbb{R}^{2M}$.

On the general grounds discussed in section \ref{subsec:geometricquantisation}, we may again define an entropic resource measure by
\begin{equation}\label{eq:SBmagic}
\begin{split}
    M_{\alpha}(|\psi \rangle) := &H_{\alpha}(p_{\psi})- M \log \pi + \frac{N}{1-\alpha}\log \alpha - N \log 2 \\
    = &\frac{1}{1-\alpha} \sum_{(a,b) \in \mathbb{F}_2^{2N}}\int_{\mathbb{R}^{2M}} \mathrm{d}\zeta \, |\chi_{\psi}(\zeta)|^2\mu \\
    &- M \log \pi + \frac{N}{1-\alpha}\log \alpha - N \log 2,
\end{split}    
\end{equation}
where we note the normalisation constant in the definition of \eqref{eq:generalprobability} is $\mu = 1/(2^N \pi^M)$. The linear shift is determined by imposing $M_{\alpha}(|\psi \rangle) = 0$ when $|\psi\rangle = |\text{Gaussian}\rangle \otimes | \text{Stabilizer} \rangle$. This is the stabilizer/Gaussian resource entropy of the combined system and automatically satisfies the properties discussed in previous section~\ref{subsubsec:entropyproperties}. 
This entropic resource measure is referred to as the {\it hybrid magic R\'enyi entropy}.
We note that, by additivity under tensor products, clearly when $|\psi\rangle = |\psi_b \rangle \otimes |\psi_s \rangle$ is a product state we have
\begin{equation}
    M_{\alpha}(|\psi \rangle) = M^S_{\alpha}(|\psi_s\rangle) + M^G_{\alpha}(|\psi_b\rangle).
\end{equation}
Finally, motivated by the analogous entropic mutual information, we define a spin-boson mutual magic measure:
\begin{equation}\label{eq:mutualentropy}
    I_{\alpha}(|\psi\rangle) := M_{\alpha}(|\psi\rangle) - M^S_{\alpha}(\rho_s) - M^G_{\alpha}(\rho_b),
\end{equation}
where $\rho_b = \tr_{\mathcal{H}_s} |\psi \rangle \langle \psi |$ and $\rho_s = \tr_{\mathcal{H}_b}  |\psi \rangle \langle \psi |$ are the partial traces over the spin and bosonic degrees of freedom respectively. The quantities $M^S_{\alpha}(\rho_s)$ and $M^G_{\alpha}(\rho_b)$ are the stabilizer and Gaussian entropies defined in equations \eqref{eq:SRE} and \eqref{eq:GRE} respectively.

In Section~\ref{sec:Examples}, we compute these quantities for a simple many-body hybrid spin-boson system: the Dicke model. We compute the resource measures introduced above using both perturbative methods and numerical simulations. We now introduce a numerical scheme to enable the latter.

\subsubsection{Numerical scheme}
The resource measure $M_{\alpha}(|\psi\rangle)$ of equation \eqref{eq:SBmagic} involves an integral over the $2M$ dimensional continuous phase space and a sum over $4^N$ Pauli strings in \eqref{eq:SBmagic}. The naive direct computation is thus exponentially costly in the system size $M+N$. In the previous works \cite{Haug_2023, Lami:2023naw, Tarabunga_2023, Tarabunga_2024}, several efficient numerical schemes have been proposed for computing the stabilizer R\'enyi entropy \eqref{eq:SRE}. Generalising these approaches to the continuous variable setting, we present in this section a Monte Carlo sampling approach to approximate $M_{\alpha}(|\psi\rangle)$.

\paragraph*{Sampling.}
We employ the Metropolis-Hastings algorithm to sample over both continuous and discrete displacement operators $\hat{D}(\xi) \otimes \sigma_{a,b}$ according to the probability distribution \( p_{\psi}(\xi, a,b) \). This probability distribution is computed pointwise on some encoding of a state $|\psi \rangle \in \mathcal{H}_b \otimes \mathcal{H}_c$ where the bosonic Fock space is truncated to maximum Boson number $N_b$. For example, later in section \ref{sec:Examples}, this state may be computed by exact diagonalisation of a local Hamiltonian $H$ using the Lanczos algorithm. The sampling scheme is summarised in Algorithm \ref{alg:metro} and we discuss the details further in the following.

In the first step of the numerical scheme, we introduce a phase space cut-off $\beta$ and sample a phase space point $\xi$ in $(\xi, a,b) \in \mathbb{R}^{2M} \times \mathbb{F}_2^{2N}$ uniformly on $[-\beta,\beta]^{2M} \times \mathbb{F}_2^{2M}$. The cut-off $\beta$ is chosen relative to the boson truncation as $N_b\approx\beta^2$. In practice, the parameter choices are verified by using the purity $\tr \rho^2 = \int \mathrm{d}\mu(\zeta) \, |\chi_{\rho}(\zeta)|^2$ as a convergence criterion.

Before collecting samples, a burn-in period $T_b$ is applied to allow the Markov chain to reach equilibrium. For each sampling step, a candidate $\xi'$ is proposed by perturbing the current $\xi$ with a Gaussian displacement of zero mean and variance $h$. For the Pauli string updates, following \cite{Lami:2023naw,Tarabunga_2023}, either a single-site update $P' = P_1 \ldots P_i' \ldots P_N$ or a two-site update $P' = P_1 \ldots P_i' \ldots P_j' \ldots P_N$ is proposed, where sites $i$ and $j$ are selected uniformly at random form the set of $N$ discrete qubit sites.

We then compute \( p(\xi', a',b') \), the proposed move is accepted with probability $\min \left(1, \frac{p(\xi', a',b')}{p(\xi, a,b)} \right).$ 
If accepted, the candidate phase space sample is updated to \((\xi', a',b')\); otherwise, the original configuration is retained. The output of the sampling algorithm is $N_{\text{Samples}}$ samples from the distribution $p_{\psi}(a,b,\xi)$ on $\mathcal{M}$ which we denote $\{\xi^{(K)},a^{(K)},b^{(K)}\}_{K=1}^{N_{\text{Samples}}}$. 

\paragraph*{Estimating.}
The quantity $M_{\alpha}(|\psi\rangle)$ may be expressed as an expectation value over $p_{\psi}(\xi,a,b)$ as follows
\begin{equation}
    e^{(1-\alpha)H_{\alpha}(|\psi\rangle)} = \mathbb{E}_{p_{\psi}}\left[p_{\psi}(\xi,a,b)^{\alpha-1}\right],
\end{equation}
and thereby estimated by 
\begin{equation}
\begin{split}
    \hat{M}_{\alpha}(|\psi\rangle) := &\frac{1}{1-\alpha}\log\mathbb{E}_{p_{\psi}}\left[p_{\psi}(\xi,a,b)^{\alpha-1}\right]\\
    &- M \log \pi + \frac{M}{1-\alpha}\log \alpha - N \log 2 .
\end{split}
\end{equation}
Following \cite{Tarabunga_2023}, for $\alpha>1$ the sample variance is given similarly by
\begin{equation}
    \mathrm{Var}(\hat{M}_{\alpha}) = \frac{\exp \left( 2(\alpha-1)(\hat{H}_{\alpha} - \hat{H}_{2\alpha - 1}) \right) - 1}{|\alpha-1|}.
\end{equation}
In addition, Monte Carlo estimates of $\hat{M_{\alpha}}$ suffer from statistical fluctuations due to sampling correlations. These are quantified via the integrated autocorrelation time \cite{Sandvik_2010, Jonathan_2010} \(\tau_I\), defined by $\tau_I = 1 + 2 \sum_{t=1}^{M} \rho(t)$ where \(\rho(t)\) is the autocorrelation function \footnote{To estimate $\rho(t)$ from a finite set of $N$ Monte Carlo samples $\{f_n\}_{n=1}^N$ with $M\ll N$, one typically uses the normalized autocorrelation estimator
\begin{equation}
\hat{\rho}_f(t) = \frac{\hat{c}_f(t)}{\hat{c}_f(0)},
\end{equation}
with
\begin{equation}
\hat{c}_f(t) = \frac{1}{N - t} \sum_{n=1}^{N - t} (f_n - \mu_f)(f_{n+\tau} - \mu_f), \quad
\mu_f = \frac{1}{N} \sum_{n=1}^N f_n.
\end{equation}
In practice, the sum is truncated at a fixed lag $M\ll N$ to avoid excessive noise from long-time correlations. We fix $M=100$.
The integrated autocorrelation time tells us how many steps are required between samples to gain statistical independence.} \cite{Sokal1997}. 
Consequently, the overall uncertainty of $\hat{M_{\alpha}}$ is therefore approximated as:
\begin{equation}
\sigma(M_{\alpha}) \approx \sqrt{\frac{\mathrm{Var}(\hat{M_{\alpha}})}{N_{\text{eff}}}}, \quad N_{\text{eff}} = {N / \tau_I}.
\end{equation}
Finally, we note a similar numerical scheme, setting $N=0$, can be used to compute the Gaussian R\'enyi entropy $M^{G}$ of equation \eqref{eq:GRE}.

\begin{algorithm}[H]
\caption{Sampling phase space $|\chi_{\psi}|^2$ distribution.}
\label{alg:metro}
\begin{algorithmic}[1]
\State \textbf{Input:} a quantum state $\lvert \psi \rangle$, number of samples $N_{\text{Samples}}$, burn-in period $T_b$, IR truncation $\beta$ and perturbation $h$.
\State Initialise a phase space point $\xi \sim \text{Unif}[-\beta,\beta]^{2M}$ and Pauli string $(a, b) \sim \text{Unif}(2^N \times 2^N)$.
\State Initialise $\texttt{Samples} = \{\}$.
\For{$i = 1$ to $T_b + N_{\text{Samples}}$}
    \State Define $(\xi',a',b')$ where $\xi' \sim \mathcal{N}(\xi,h)$ and $(a',b')$ are uniform random single or double site binary flip of $(a,b)$.
    \State Set $q \gets \min\left(1, \frac{p(\xi', a',b')}{p(\xi,a,b)} \right)$ and $u \gets \text{Unif}[0,1]$.
    \If{$u < q$}
    \State $(\xi,a,b) \gets (\xi',a',b')$.
    \EndIf
    \If{$i > T_b$}
    \State \text{append $(\xi,a,b)$}  \text{ to }    \texttt{Samples}
    \EndIf
\EndFor
\State \textbf{Return:} \texttt{Samples} a set of $N_\text{Samples}$ points $(\xi, a, b)$.
\end{algorithmic}
\end{algorithm}

\section{Examples}\label{sec:Examples}
In this section, we compute the magic resource entropy of a number of many-body examples. While the entanglement properties of many-body ground states have been extensively studied, there has been recent growing interest in understanding their stabilizer properties. In particular, recent work has highlighted the ability of magic to detect and distinguish quantum phases.

Several studies have investigated stabilizer resources in this context. For example, the stabilizer R\'enyi entropy of the transverse-field Ising model was analysed in~\cite{oliviero2022magic}, while the Mana of the three-state Potts model was considered in~\cite{tarabunga2024critical}, where universal scaling behaviours were identified at critical points. More recently,~\cite{hoshino2025stabilizer} provided a physical interpretation of these universal contributions in terms of defect partition functions in conformal field theory (CFT), identifying the sub-leading universal term in the stabilizer R\'enyi entropy with a boundary $g$-factor.

We begin by studying the Gaussian R\'enyi entropy~\eqref{eq:GRE} of the ground states of a number of elementary bosonic models. We then turn to our main example: the Dicke model, where we compute the hybrid spin-boson resource measure~\eqref{eq:SBmagic} both perturbatively and numerically. Finally, we examine the dynamics of these resource entropic quantities following a quench in the Jaynes–Cummings model.

\subsection{Gaussian entropy}

\subsubsection{Fock states}
\label{sec:Fock}
We begin with an illustrative exercise comparing the Gaussian R\'enyi entropies of the single boson Fock states $|0 \rangle$, $|1\rangle$ and $|\pm \rangle := (|0\rangle \pm |1 \rangle)/ \sqrt{2}$.

In this section we work with real coordinates $\xi = (\xi_q,\xi_p)$ on the single boson phase space $\mathcal{M} = T^*\mathbb{R}$. We denote the Weyl function of a state $|\psi \rangle \in L^2(\mathbb{R})$ by $\chi_{\psi}(\xi_{p},\xi_{q}) = \bra{\psi}\hat{D}(\xi_q,\xi_p)\ket{\psi}$ with the displacement operators as defined in equation \eqref{eq:CVdisplacement}. 

The Weyl function of the Gaussian state $|0\rangle$ is $\chi_{\ket{0}}(\xi_{p},\xi_{q}) = e^{-r^2/4}$, where $r^2 = \xi_q^2 + \xi_p^2$, with corresponding probability density function \eqref{eq:generalprobability} given by
\begin{align}
  p_{\ket{0}}(\xi_{p},\xi_{q}) = \frac{1}{2\pi}e^{-r^2/2},
\end{align}
The linear shifts of the Gaussian R\'enyi entropy definition \eqref{eq:GRE} are chosen such that $M^G_{\alpha}(|0\rangle) =0$.

We now consider the boson number eigenstate $\ket{1}$. The Weyl function is given by
\begin{equation}
\chi_{\ket{1}}(\xi_q,\xi_p) = e^{-r^2/4}L_{1}\left(\frac{r^2}{2}\right),
\end{equation}
where $L_n(x)$ denotes the $n^{\text{th}}$ Laguerre polynomial. The corresponding probability density function is
\begin{align}
p_{\ket{1}}(\xi_{p},\xi_{q}) = \frac{1}{2\pi}e^{-r^2/2}\left( 1 - \frac{r^2}{2}\right)^{2},
\end{align}
and the Gaussian differential entropy is
\begin{equation}
\begin{split}
M^G(\ket{1}) &= \int_{\mathbb{R}^2} \frac{\mathrm{d}\xi_p\, \mathrm{d}\xi_q}{2\pi} \, 
e^{-r^2/2} \left( 1 - \frac{r^2}{2} \right)^2 \\
&\quad \times \left( \frac{r^2}{2} - 2 \log \left| 1 - \frac{r^2}{2} \right| \right) \\
&= 2 \int_{0}^{\infty}  \mathrm{d}r\, r e^{-r^2} (1 - r^2)^2 
\left( r^2 - 2 \log|1 - r^2| \right) \\
&\simeq 2.3943\ldots
\end{split}
\end{equation}
Finally, we consider the states $|\pm \rangle$. The Weyl functions are given by
\begin{equation}
  \chi_{|\pm\rangle}(\xi_{q},\xi_{p}) =  e^{-\frac{1}{4}(\xi_{p}^{2} + \xi_{q}^{2})}\left( 1 - \frac{1}{4}(\xi_{p}^{2} + \xi_{q}^{2}) \pm \frac{i}{\sqrt{2}}\xi_{p} \right),
\end{equation} 
with corresponding probability density functions
\begin{align}
p_{|\pm \rangle}(\xi_q,\xi_p) = \frac{1}{2\pi}e^{-r^2/2} \left(1 -\frac{r^2}{2} +\frac{r^4}{16} + \frac{1}{2}\xi_{p}^{2}\right).
\end{align}
The Gaussian differential entropy is equal for both states and given by the integral
\begin{equation}
\begin{split}
M^G(|\pm \rangle) &= 2 \int_0^{\infty}  \mathrm{d}r \mathrm{d}\theta\, r p_{|\pm \rangle}\, \bigg( 
r^{2} \\
&\quad - \log \left| 1 - r^{2} + \tfrac{1}{4} r^{4} + r^{2} \sin^{2}(\theta) \right| \bigg) \\
&\simeq 1.2775\ldots
\end{split}
\end{equation}
We observe that $0 = M^G(\ket{0}) < M^G(\ket{\pm}) < M^G(\ket{1})$.

\subsubsection{Anharmonic oscillator}
We now consider the anharmonic oscillator Hamiltonian
\begin{align}
  H = \frac{\hat{p}^{2}}{2m} + \frac{\hat{x}^2}{2} +  \frac{\lambda}{4}{\hat{x}}^4 = H_0 + \frac{\lambda}{4}{\hat{x}}^4,
\end{align}
and study the perturbation theory over the ground state $\ket{0}$ of the harmonic oscillator $H_{0}$. The (unnormalised) first-order correction to the ground state is
\begin{align}
  \begin{split}
  \ket{\Omega} &= \ket{0} + \frac{\lambda}{4} \sum_{n > 0}\frac{\bra{n}\hat{x}^4\ket{0}}{E_{n} - E_{0}}\ket{n} + \mathcal{O}(\lambda^{2}) \\
   &= \ket{0} -\lambda \bigg( \frac{3}{4}(a^{\dagger})^{2}\ket{0} + \frac{1}{16}(a^{\dagger})^{4}\ket{0} \bigg) + \mathcal{O}(\lambda^{2}),
  \end{split}
\end{align}
where $E_n$ denote the harmonic oscillator energy levels. Working in complex coordinates $z = \xi_q + i \xi_p = r e^{i \theta}$ on phase space, we similarly expand the Weyl function
\begin{align}
  \chi_{|\Omega\rangle}(z) = \chi_{|\Omega \rangle}^{(0)}(z) + \lambda \chi_{|\Omega \rangle}^{(1)}(z) + \lambda^{2} \chi_{|\Omega\rangle}^{(2)}(z)  + \mathcal{O}(\lambda^{3}),
\end{align}
here we consider perturbation theory to quadratic order since the stabilizer content of the ground state is invariant under $\lambda \to -\lambda$ and the non-trivial contribution begins at quadratic order. The leading term in the Weyl function expansion is given by the Gaussian ground state  
$\chi_{|\Omega \rangle}^{(0)}(z) = e^{-r^{2}/2}$, and the corrections are found to be
\begin{align}
\begin{split}
\chi_{|\Omega \rangle}^{(1)}(z) 
&= -e^{-r^{2}/2} \bigg( 
\frac{3}{2} r^{2} \cos 2\theta 
+ \frac{1}{8} r^{4} \cos 4\theta 
\bigg),
\end{split} \\
\begin{split}
\chi_{|\Omega \rangle}^{(2)}(z) 
&= -e^{-r^{2}/2} \bigg( 
\frac{9}{4} L_2(|z|^2) 
+ \frac{9}{4} L_4(|z|^2) \\
&\quad 
+ \frac{3}{32} r^{2}  
\left(12 - 8r^{2} + r^{4} \right) \cos 2\theta
\bigg).
\end{split}
\end{align}
The (unnormalised) probability density function \eqref{eq:generalprobability} is then given perturbatively by
\begin{align}
\begin{split}
\tilde{p}_{|\Omega\rangle}(z) 
&= \frac{1}{\pi} \bigg( 
|\chi_{|\Omega \rangle}^{(0)}(z)|^{2} 
+ 2 \lambda\chi_{|\Omega \rangle}^{(0)}(z) \chi_{|\Omega \rangle}^{(1)}(z)  \\
&\quad + \lambda^{2} \left( 
|\chi_{|\Omega \rangle}^{(1)}(z)|^{2} 
+ 2 \chi_{|\Omega \rangle}^{(0)}(z) \chi_{|\Omega \rangle}^{(2)}(z) 
\right)  
\bigg) 
+ \mathcal{O}(\lambda^{3}).
\end{split}
\end{align}
To account for the normalisation of the ground state $|\Omega\rangle$, we compute to quadratic order
\begin{align}
\begin{split}
&\int_{\mathcal{M}} \mathrm{d}^{2}z\, 
|\tilde{p}_{|\Omega\rangle}(z)|^{2} 
= \left( 1 + \frac{147}{128} \lambda^{2} \right)^{-4} 
\int_{\mathcal{M}} \frac{\mathrm{d}^{2}z}{\pi^{2}} \bigg[ 
|\chi_{|\Omega \rangle}^{(0)}(z)|^{4} \\
&\quad \quad\bigg( 
6 \big| \chi_{|\Omega \rangle}^{(0)}(z) 
\chi_{|\Omega \rangle}^{(1)}(z) \big|^{2}
+ 4 \big| \chi_{|\Omega \rangle}^{(0)}(z) \big|^{3} 
\chi_{|\Omega \rangle}^{(2)}(z) 
\bigg) \lambda^{2} 
\bigg] \\
&\quad \quad= \left( 1 + \frac{147}{128} \lambda^{2} \right)^{-4} 
\left( \frac{1}{2\pi} + \frac{81}{256\pi} \lambda^{2} \right),
\end{split}
\end{align}
and perturbatively normalise the probability distribution according to $p_{|\Omega\rangle}(z) := (1 + \frac{147}{128}\lambda^{2})^{2} \tilde{p}_{|\Omega\rangle}(z)$. Finally, we find that the Gaussian $2$-R\'enyi entropy to quadratic order is given by
\begin{align}
\begin{split}
&M^{G}_2(\ket{\Omega}) 
= -\log \left( \int_{\mathcal{M}} \mathrm{d}^{2}z \, |p_{|\Omega\rangle}(z)|^{2} \right) - \log 2\pi\\
&\quad \simeq  
- \log \left( 1 + \frac{81}{128} \lambda^{2} \right) 
+ 4 \log \left( 1 + \frac{147}{128} \lambda^{2} \right) + \mathcal{O}(\lambda^2) \\
&\quad \simeq  \frac{507}{128} \lambda^{2} 
> M_{2}^G(\ket{0}).
\end{split}
\end{align}
The value is larger than the Gaussian harmonic oscillator ground state as expected.

\subsubsection{Bose-Hubbard model}
We now discuss the ground states of the Bose-Hubbard model,
\begin{equation}
    H := -J\sum_{i=1}^L(b^{\dagger}_{i}b_{i+1} + b_{i+1}^{\dagger}b_{i}) + \mu \sum_{i=1}^L n_{i} + \frac{U}{2}\sum_{i=1}^L n_{i}(n_{i}  -1),
\end{equation}
in the limits corresponding to the deep superfluid and Mott-insulating phases. In the superfluid phase $J \gg U$, the ground state is \cite{capello2007superfluid}
\begin{equation}
    \ket{\Omega_{\text{SF}}} = \prod_{i = 1}^{L}e^{\sqrt{N/L}b^{\dagger}_{i}}\ket{0},
\end{equation}
which is expressed in terms of the fixed average number density $\langle b^{\dagger}_{i}b_{i}\rangle = N/L$. The ground state is a coherent state and so the Gaussian R\'enyi entropy in the superfluid phases scales as $M^{G}(\ket{\Omega_\text{SF}}) \sim \mathcal{O}(1)$ in the system size. On the other hand, deep in the Mott insulating phase, $U \gg J$, the bosons are confined by the on-site interaction and the many-body ground state is
\begin{equation}
    \ket{\Omega_\text{MI}} = \prod_{i = 1}^{L}\frac{1}{\sqrt{n_{0}}}(b^{\dagger}_{i})^{n_{0}}\ket{0},
\end{equation}
where $n_{0}$ denotes the local occupation number for each lattice site. For $\mu/U> 0$ the local occupation is non-zero and $\ket{\Omega_\text{MI}} = \ket{n_{0}}^{\otimes L}$. Thus, by the additivity under tensor products of the Gaussian R\'enyi entropy we have the scaling $M^{G}(\ket{\Omega_\text{MI}}) \sim L$, where the proportionality constant is fixed by the entropy of $\ket{n_{0}}$.

\subsection{Hybrid magic R\'enyi entropy}
The main example we consider to illustrate hybrid magic R\'enyi entropy is the Dicke model \cite{dicke1954coherence}. The Dicke model is a model of a single photon mode coupled to $N$ spins. The Hilbert space is then $\mathcal{H} = \mathcal{H}_b \otimes \mathcal{H}_s$ with $\mathcal{H}_b = L^2(\mathbb{R})$, a one mode CV system, and $\mathcal{H}_s = (\mathbb{C}^2)^{\otimes N}$, a system of $N$ qubits. In this section and the following we label the basis states of $\mathbb{C}^2$ as $\ket{\uparrow}$ and $\ket{\downarrow}$. The Dicke model Hamiltonian reads
\begin{equation}\label{eq:DickeHamiltonian}
    H := \omega_c \hat{a}^{\dag} \hat{a} + \omega_z \sum_{j=1}^N Z_j + \frac{2\lambda}{\sqrt{N}}(\hat{a}+\hat{a}^\dag) \otimes \sum_{j=1}^N X_j.
\end{equation}
It is well known that at large $N$ the model undergoes a quantum phase transition at $\lambda = \lambda_c = \sqrt{\omega_c \omega_z}/2$. There are two phases, namely \textit{normal} and \textit{superradiant}. There is a $\mathbb{Z}_2$ parity symmetry that is spontaneously broken in the superradiant phase with an associated order parameter $\langle a \rangle / \sqrt{N}$. 

The Dicke model phase transition has previously been studied from the perspective of `entanglement resource theory' by Lambert \textit{et al.}~\cite{lambert2004entanglement,lambert2005entanglement}, where the critical point was identified via a divergence in the von Neumann entanglement entropy. In this work, we revisit the transition through the lens of stabilizer resource theory. We compute the Gaussian R\'enyi entropy~\eqref{eq:GRE}, the stabilizer R\'enyi entropy~\eqref{eq:SRE}, the hybrid magic R\'enyi entropy, and the mutual magic entropy~\eqref{eq:mutualentropy} both numerically and perturbatively. We use the Dicke model as a prototypical example of a spin-boson system; the computational methods we develop are presented generally and we do not rely on special properties of the Dicke model (such as total spin variables or the Holstein-Primakoff transformation).

\subsubsection{Numerical results}\label{subsubsec:DickeNumerics}
We now apply the numerical scheme introduced in section \ref{subsec:spinbosonmagic}. The Dicke model Hamiltonian \eqref{eq:DickeHamiltonian} has a non-degenerate ground state in the normal phase and in the superradiant phase we select the even parity ground state. The input quantum state $|\Omega \rangle$ to algorithm \ref{alg:metro} is then given by the ground state obtained by exact diagonalisation (using the Lanzcos algorithm) in the truncated Hilbert space with basis $|n\rangle \otimes |x\rangle$ where $n=1,2,\ldots N_b$ and $|x\rangle$ is a spin basis state. In the following, $\rho_b$ and $\rho_s$ denote the reduced density matrices of the boson and spin degrees of freedom respectively.

\begin{figure*}
    \begin{subfigure}[b]{0.24\textwidth}
        \includegraphics[width=\textwidth]{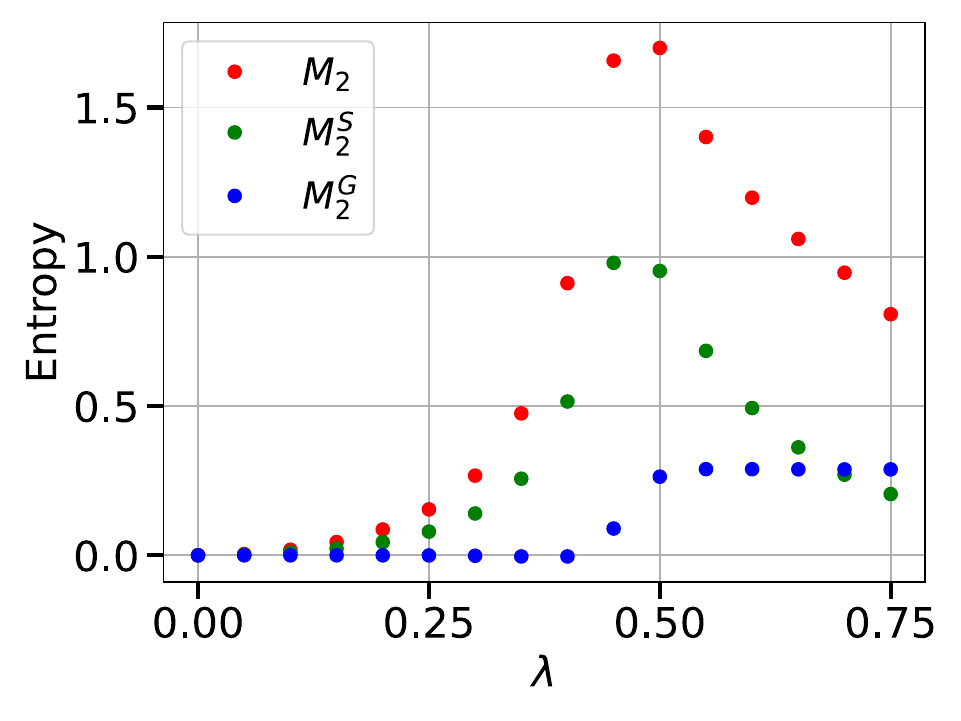}
        \caption{}
    \end{subfigure}
    \begin{subfigure}[b]{0.24\textwidth}
        \includegraphics[width=\textwidth]{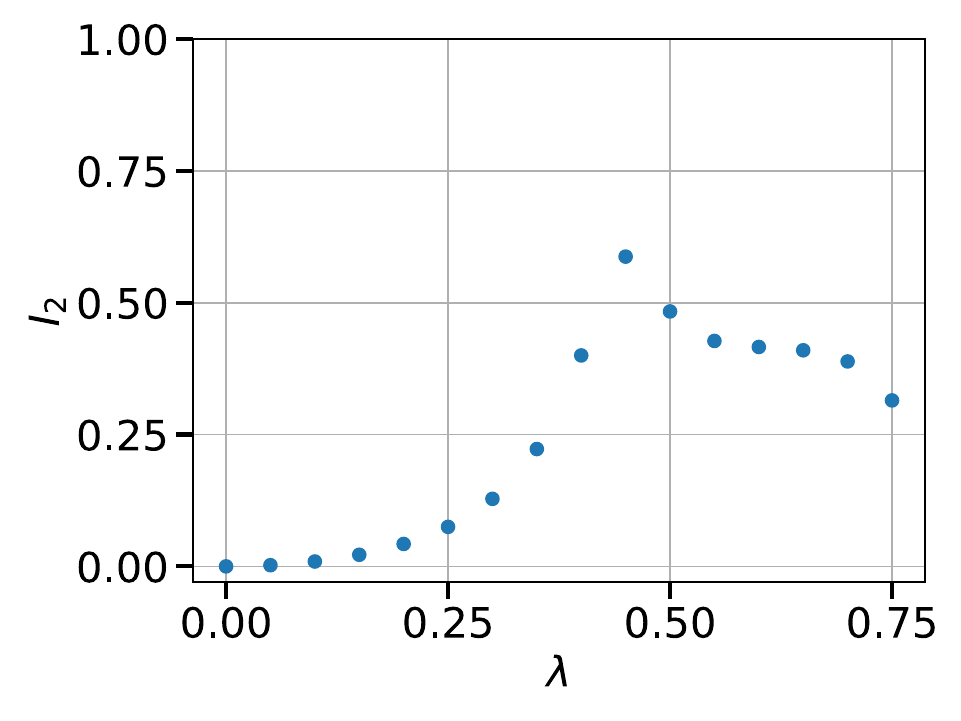}
        \caption{}
    \end{subfigure}
    \begin{subfigure}[b]{0.24\textwidth}
        \includegraphics[width=\linewidth]{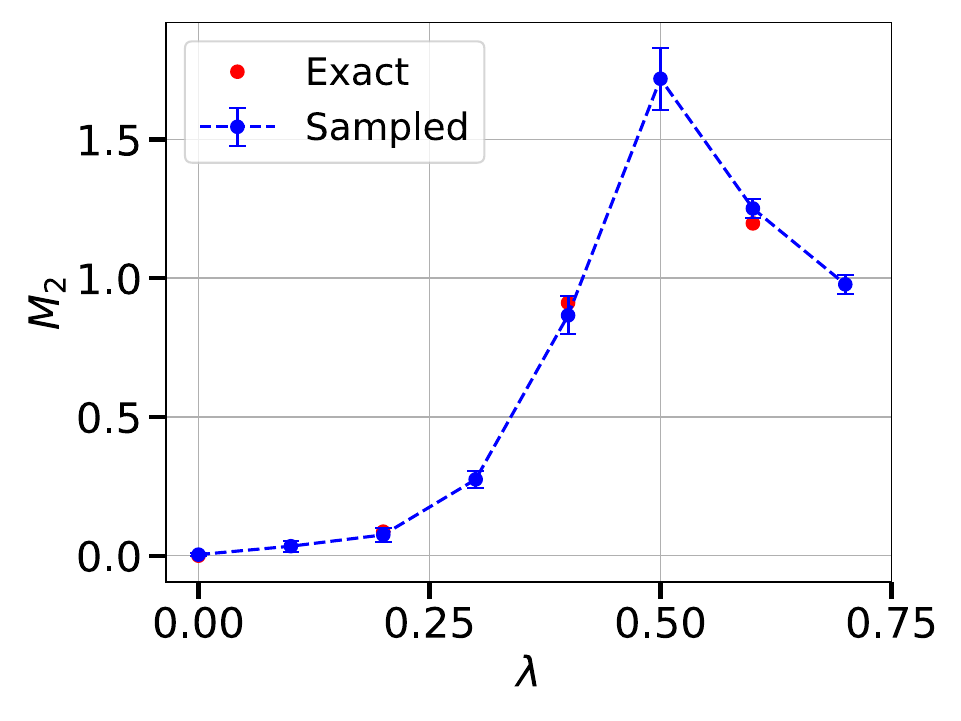}
        \caption{}
    \end{subfigure}
    \begin{subfigure}[b]{0.24\textwidth}
        \includegraphics[width=\linewidth]{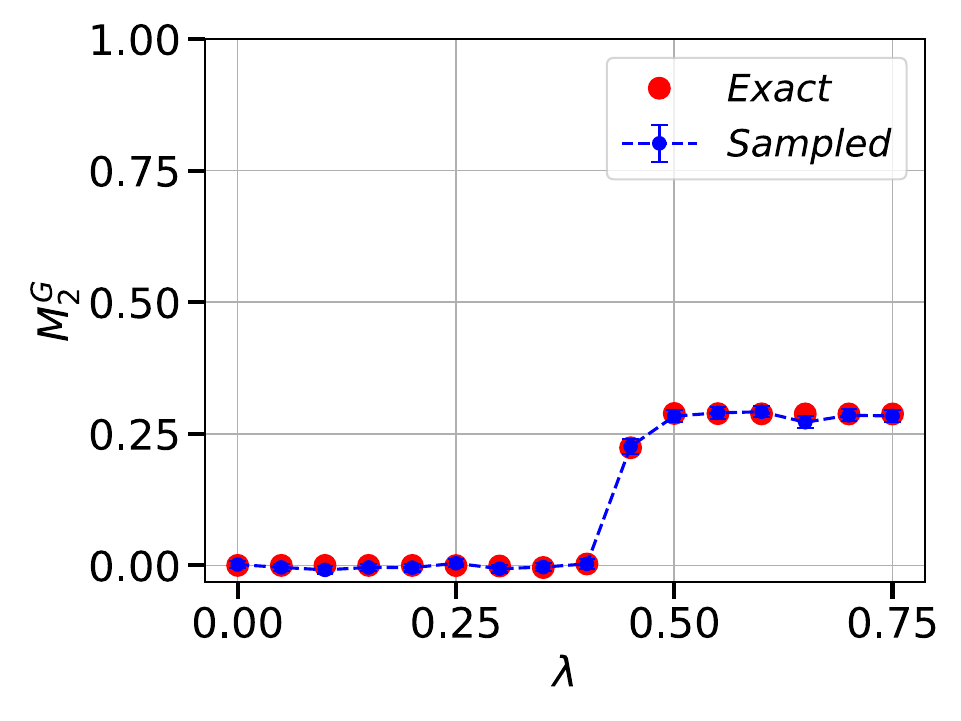}
        \caption{}
    \end{subfigure}
    \caption{Subfigures (a) and (b) show the exact diagonalisation calculations for $M_2$, $M^S_2$, $M^G_2$ and the mutual measure $I_2$ respectively with \( N_{\text{Spin}} = 4 \) using truncations \( N_b = 150 \) and \(\beta = 12\). (c) Comparison between exact diagonalisations and sampling results for spin-boson magic entropy $M_2$. We take parameters \( N_b = \beta^2 \) with \(\beta = 14\), and the deviation parameter \( h = 0.1 \beta \). (d) Monte Carlo results for Gaussian R\'enyi entropy. $N_{\text{Spin}}=6$, truncation $\beta=12$ and $N_{\text{Samples}}=10^5$ with deviation $h=1.2$.}
    \label{fig:Dickeresults}
\end{figure*}

The numerical calculations for $M_{\alpha}^G(\rho_b)$, $M_{\alpha}^S(\rho_s)$, and $M_{\alpha}(|\psi\rangle)$ are shown in Figure \ref{fig:Dickeresults}. In the numerical calculation we set $\omega_c=\omega_z=1$ and consider the case of $N=4$. We compute the various resource measures with $\alpha=2$. We select the Monte Carlo sampling parameters  $T_b = 1000$, $\beta=14$, $N_b=\beta^2$, $h=0.1\beta$ and $N_{\text{Samples}} = 10^5$. At low system size we are able to benchmark the results against an exact numerical integration approach.

Our numerical results suggest that at the superradiant critical point $\lambda = 1/2$ the stabilizer R\'enyi entropy $M^S_{2}(\rho_s)$ and hybrid magic R\'enyi entropy $M_2(|\psi \rangle)$ diverge whereas the Gaussian R\'enyi entropy $M^G_{2}(\rho_b)$ shows a step-like transition. When $\lambda = 0$ these are quantities are zero, as expected for a state of the form $|\psi \rangle = |\text{Gaussian}\rangle \otimes |\text{Stabilizer}\rangle$. Figure \ref{fig:Dickeresults} also includes a plot of the mutual measure $I_2(|\psi\rangle)$. We see that bosonic non-Gaussianity and spin non-stabilizerness are generated as $\lambda$ is increased. At the critical point the mutual \eqref{eq:mutualentropy} is maximal indicating that the resource is maximally spread between the boson and spin degrees of freedom.

\begin{figure}
    \centering
    \includegraphics[scale=0.4]{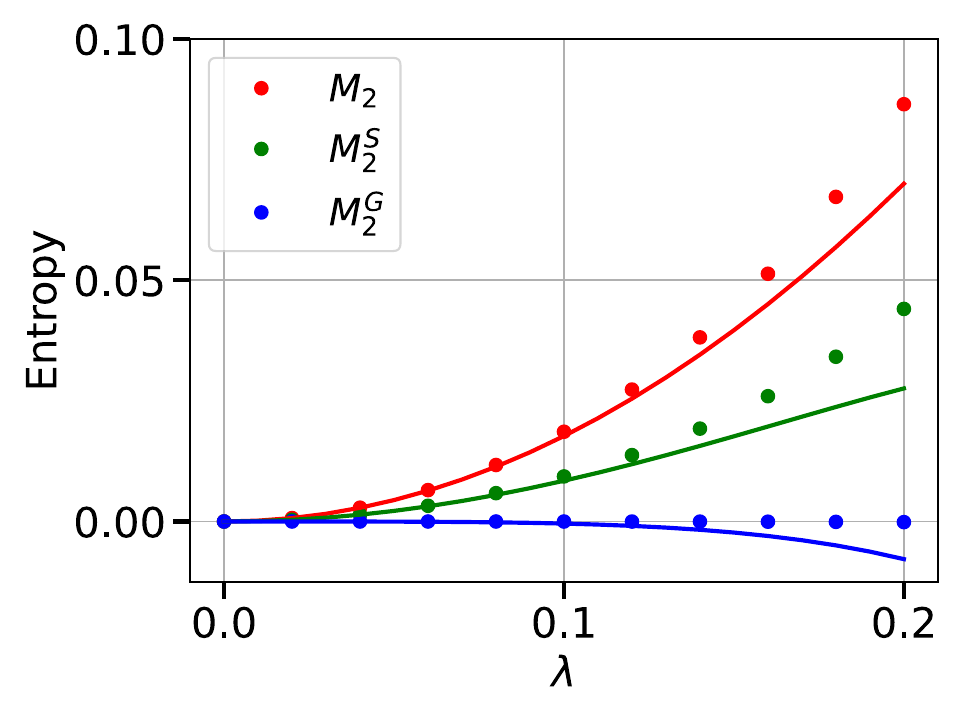}
    \caption{Comparison between numerical and perturbative results for $M_2$, $M^S_2$ and $M^G_2$ in the small-$\lambda$ regime. Solid lines indicate perturbative predictions and the numerical data is as in figure \ref{fig:Dickeresults}.}
    \label{fig:perturbative}
\end{figure}

\subsubsection{Perturbative calculations.}
The ground state $|\Omega \rangle$ of the Dicke model Hamiltonian \eqref{eq:DickeHamiltonian} at $\lambda=0$ is a product of a Gaussian and stabilizer state. In this section we compute the perturbative correction to the ground state and the spin-boson entropy $M_2(|\Omega \rangle)$; the Gaussian R\'enyi entropy $M^{G}_2(|\Omega \rangle)$; and the stabilizer R\'enyi entropy $M^S_2(|\Omega\rangle)$ to first order in the coupling constant $\lambda$. We state the results in the following and refer the reader to appendix \ref{appendix:Dicke} for further details.

We find the first order correction to the ground state of the Dicke model is given by

\begin{equation}
  \ket{\Omega} = \frac{1}{\sqrt{\mathcal{N}}}\bigg(\ket{0}\ket{\downarrow} - \sqrt{\frac{\kappa}{N}}\sum_{i=1}^N X_{i}\ket{1}\ket{\downarrow} \bigg) +   \mathcal{O}(\lambda^2),
\end{equation}
where $\sqrt{\kappa} = 2\lambda / (\omega_{c} + 2\omega_{z})$ and, to first order in $\lambda$, the normalisation constant is $\mathcal{N} = 1 + \kappa$. We then compute the associated perturbative Weyl function and find the following result for the hybrid magic R\'enyi entropy with $\alpha=2$:
\begin{align}
  \begin{split}
  &M_{2}(|\Omega \rangle)
  = 4\log(\mathcal{N}) \\
    & \quad - \log\bigg( 1 +  \frac{6\kappa^{2}}{N} + \frac{1}{2}\frac{(N^{2} - 2N + 2)\kappa^{4}}{N^{3}} \bigg) + \mathcal{O}(\lambda^2).
  \end{split}
\end{align}
The Gaussian R\'enyi entropy calculation requires us to trace out the spin degrees of freedom. We find
\begin{align}
  \rho_{b} = \tr_{\mathcal{H}_s} \rho = \mathcal{N}^{-1}(\ketbra{0}{0} + \kappa \ketbra{1}{1}) + \mathcal{O}(\lambda^2),
\end{align}
and the corresponding Gaussian R\'enyi entropy is 
\begin{align}
  \begin{split}
  &M_{2}^G(\rho_{b}) 
  = \log \tr \rho_{b}^{2} + 4\log \mathcal{N} \\
  & \quad - \log(1 + 2\kappa + 3\kappa^{2} + \kappa^{3} + \frac{1}{2}\kappa^{4}) + \mathcal{O}(\lambda^2).
  \end{split}
\end{align}
Similarly, tracing over the boson Hilbert spaces gives
\begin{align}
  \rho_{s} = \mathcal{N}^{-1}\left( \ketbra{\downarrow}{\downarrow} + \frac{\kappa}{N}\sum_{i,j=1}^N X_{i}\ketbra{\downarrow}{\downarrow}X_{j} \right) + \mathcal{O}(\lambda^2),
\end{align}
and we find the corresponding stabiliser R\'enyi entropy is 
\begin{align}
  \begin{split}
  &M_{2}^S(\rho_{s}) 
  = \log \tr \rho_{s}^{2} + 4\log \mathcal{N} \\  
  & \quad - \log\left( 1 + \frac{6\kappa^{2}}{N} + \frac{(N^{2} - 2N + 2)\kappa^{4}}{N^{3}} \right) + \mathcal{O}(\lambda^2).
  \end{split}
\end{align}
We observe that, to first order in $\lambda$, the inequality $M_2(|\Omega\rangle) \ge M_2^G(\rho_b) + M_2^{S}(\rho_s)$ holds. This implies that the mutual magic entropy~\eqref{eq:mutualentropy} is positive for small $\lambda$, and vanishes at the non-interacting point $\lambda = 0$. As demonstrated in Figure~\ref{fig:perturbative}, the perturbative calculation agrees well with the numerical results presented in Section~\ref{subsubsec:DickeNumerics} for small values of $\lambda$. However, the first-order ground state has a boson occupation number equal to one, and the perturbative expansion breaks down as $\lambda$ approaches the critical point, where the boson number diverges.

\subsubsection{Quench dynamics in the Jaynes-Cummings model}

To investigate the spreading of non-stabilizer and non-Gaussian quantum resources under non-equilibrium dynamics, we now turn to a minimal example: the Jaynes–Cummings model~\cite{jaynes2005comparison}. Physically, this system models the coherent interaction between a single two-level atom and a single mode of the quantised electromagnetic field. The Hilbert space is then $\mathcal{H} = L^2(\mathbb{R}) \otimes \mathbb{C}^2$ and the Hamiltonian is given by
\begin{equation}
\label{eq:JC_model}
H = \omega_c \hat{a}^\dagger \hat{a} + \frac{1}{2}\omega_z Z + \lambda \left( \hat{a} \otimes \sigma_+ + \hat{a}^{\dagger} \otimes \sigma_- \right),
\end{equation}
where $\sigma_{\pm} = (X \pm i Y)/2$. The parameters $\omega_c$, $\omega_z$ and $\lambda$ denote the cavity photon frequency, the energy splitting of the two-level atom, and the atom-field coupling strength respectively. The first two terms of the Hamiltonian represent the free dynamics of the field and atom, respectively, and the interaction term conserves total excitation number: the atom absorbs a photon to transition to the excited state and emits a photon to return to the ground state.

We study the time evolution of the following resource measures: the Gaussian R\'enyi entropy~\eqref{eq:GRE}, the stabilizer R\'enyi entropy~\eqref{eq:SRE}, the hybrid magic R\'enyi entropy~\eqref{eq:SBmagic}, and the mutual magic entropy~\eqref{eq:mutualentropy} using the numerical scheme introduced previously in section \ref{subsec:spinbosonmagic}.
We consider the time evolution of two initial states namely $|n\rangle \ket{\uparrow}$ and $|\gamma \rangle\ket{\uparrow}$ for a coherent state labelled by $\gamma \in \mathbb{C}$. We consider the resonance condition $\omega_c=\omega_z=1$ and consider the coupling $\lambda=1$ throughout.

\begin{figure*}
    \centering
    \begin{subfigure}[b]{0.24\textwidth}
        \includegraphics[width=\textwidth]{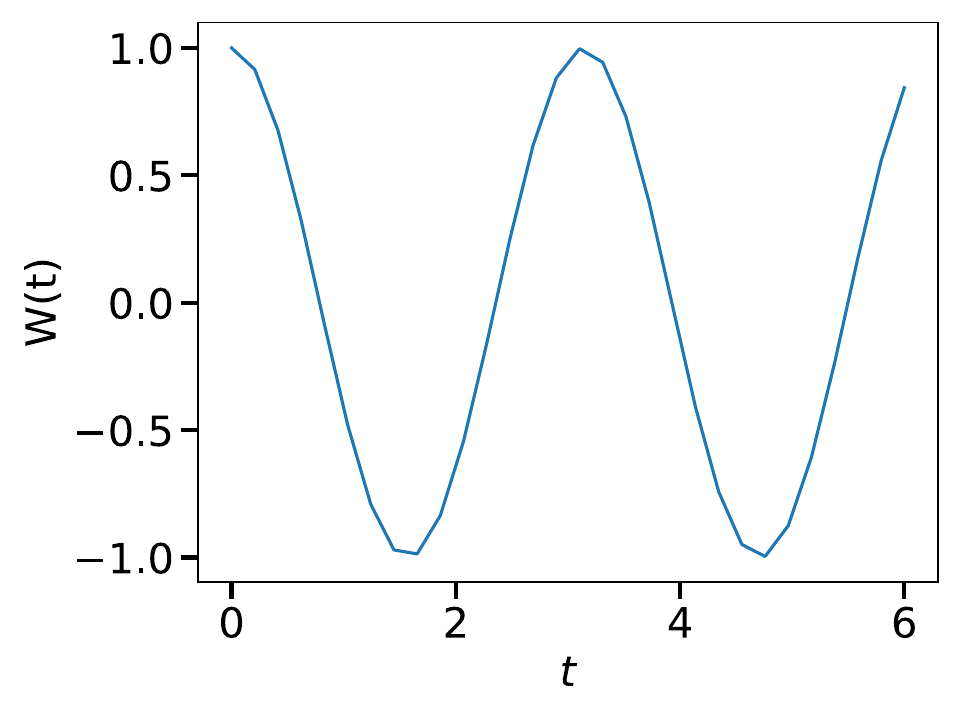}
        \caption{}
    \end{subfigure}
    \begin{subfigure}[b]{0.24\textwidth}
        \includegraphics[width=\textwidth]{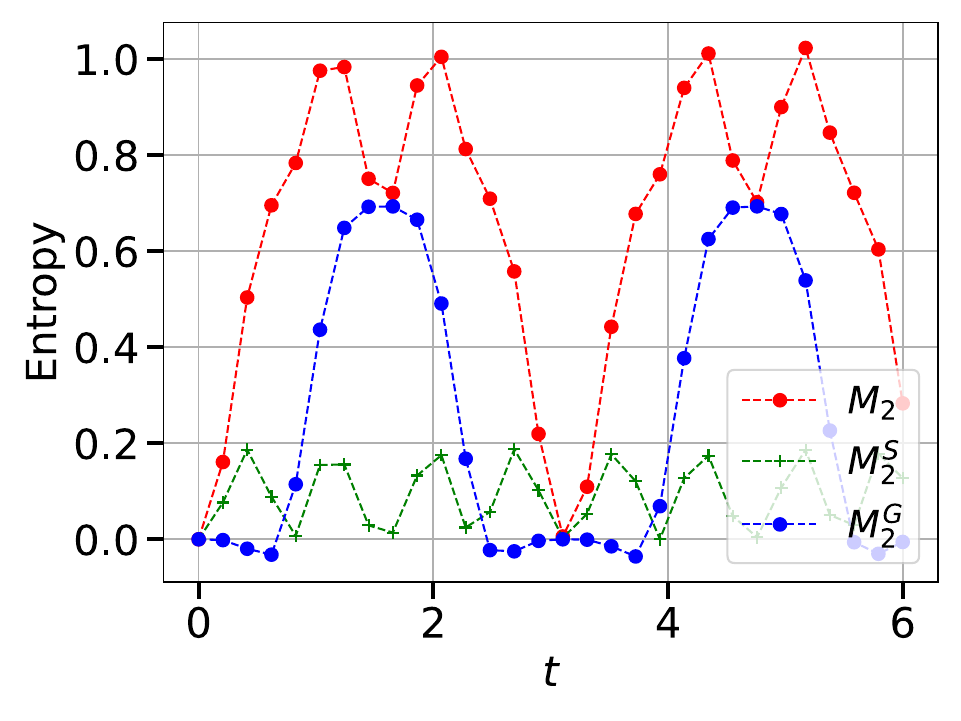}
        \caption{}
    \end{subfigure}
    \begin{subfigure}[b]{0.24\textwidth}
        \includegraphics[width=\textwidth]{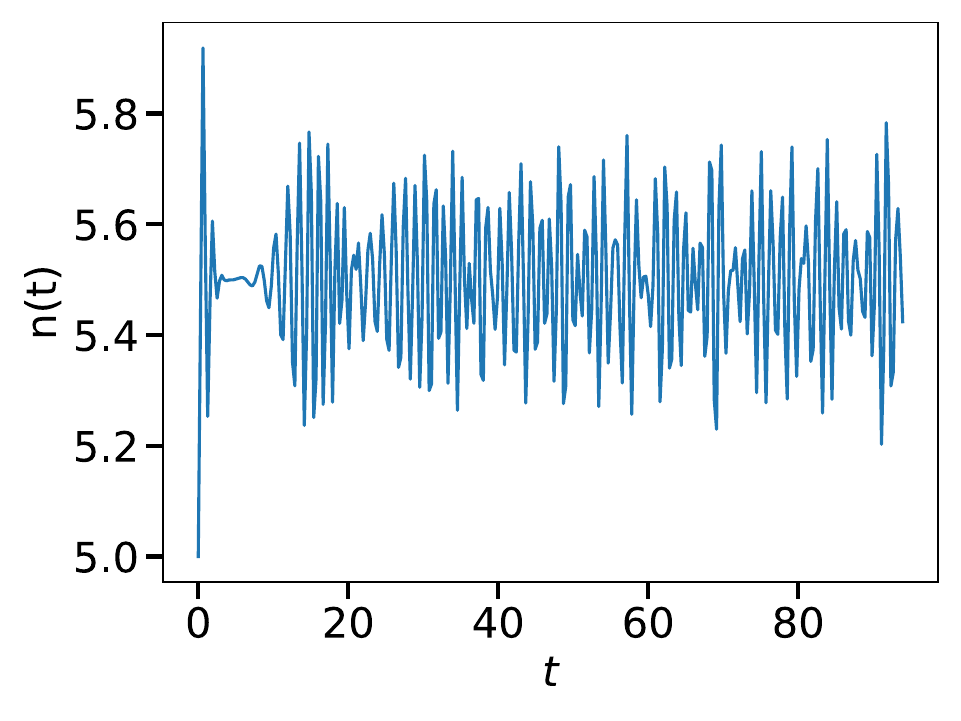}
        \caption{}
    \end{subfigure}
    \hfill
    \begin{subfigure}[b]{0.24\textwidth}
        \includegraphics[width=\textwidth]{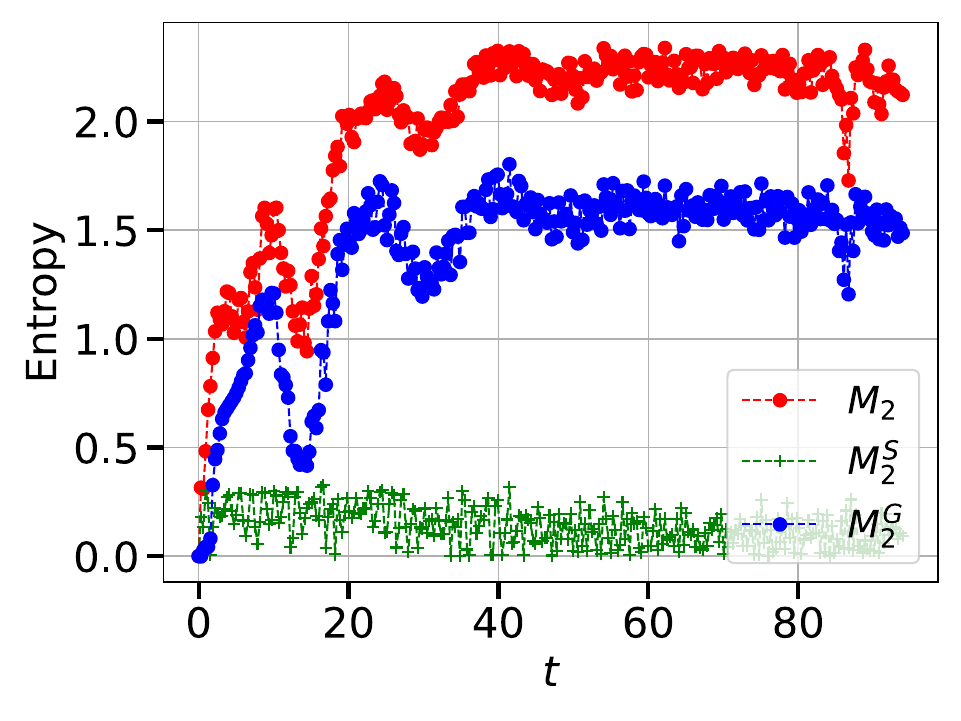}
        \caption{}
    \end{subfigure}    
    \caption{Rabi oscillation in Jaynes-Cumming model with initial state $|0,e\rangle$. (a) Atomic inversion dynamics. (b) Dynamics of the resource entropy measures $M^G_2$ and $M_2$. Sampling parameters are $\beta = 10$ with $N_b=\beta^2$.   Jaynes–Cummings model quench dynamics with a coherent state $\gamma=5$ in the photon sector. (c) Photon number dynamics. (d) Spin-boson entropy, stabilizer R\'enyi entropy, and Gaussian entropies, calculated by exact diagonalisation. The spin-boson and Gaussian entropy calculations are considered with bosonic truncations \( N_b = \beta^2 \) and \(\beta = 15\).}
    \label{fig:JC_dynamics_n0}
\end{figure*}

Let us first consider the initial state $|\psi(0)\rangle=|n\rangle\ket{\uparrow}$. The state will undergo Rabi oscillation \cite{Gerry_Knight_2004} between $|n\rangle\ket{\uparrow}$ and $|n+1\rangle\ket{\downarrow}$ due to the interaction term. To detect these excited state and ground state populations in the spin sector, we consider the atomic inversion defined by $W(t) = \langle\psi(t)| Z |\psi(t)\rangle$. The figure \ref{fig:JC_dynamics_n0}(a) shows that $W(t)$ oscillates between $|0\rangle\ket{\uparrow}$ and $|1\rangle \ket{\downarrow}$ with the quantum electrodynamic Rabi oscillation period $T = \pi$. All the quantum resource entropies similarly oscillate in time and are plotted in figure \ref{fig:JC_dynamics_n0}(b). We find $M_2 = M^G_2 + M^S_2$ at $t=0$ and $t=\pi$ as expected from a product state. In addition (and as expected from the discussion in section \ref{sec:Fock}), the Gaussian R\'enyi entropy increases as $n$ increases when state evolves from $|0\rangle\ket{\uparrow}$ to $|1\rangle\ket{\downarrow}$. The time evolution of the stabilizer R\'enyi entropy oscillates more rapidly and has a smaller amplitude since for a single spin $M^S_2$ is bounded \footnote{The bound is straightforward to compute by substituting the three variable Weyl function for $\rho$ into the definition of stabilizer R\'enyi entropy and maximising.} by $\log(3/2)$ for single qubit mixed states.  

Let us now consider the initial state with a coherent state $|\gamma \rangle = \sum_{n=0}^{\infty}{e^{-|\gamma|^2/2}\frac{\gamma^n}{\sqrt{n!}}|n\rangle}$ in the photon sector. We compute the photon number $n=a^{\dagger}a$ in figure \ref{fig:JC_dynamics_n0}(c) and observe a sequence of collapses and revivals, the revivals becoming less distinct as time increases. In figure \ref{fig:JC_dynamics_n0}(d) we plot the various quantum resource measures. We note that the stabilizer and Gaussian entropies exhibit different behaviour. Both the Gaussian and the hybrid magic R\'enyi entropies increase in time (due to the de-phasing of the coherent state under the quench dynamics) whereas the stabilizer R\'enyi entropy is bounded.

\section{Discussion}\label{sec:discussion}
We have introduced a stabiliser/Gaussian resource measure that extends the stabiliser R\'enyi entropy to hybrid spin-boson systems. By highlighting the shared structure of stabilizer and Gaussian R\'enyi entropies from a phase space perspective, we proposed a natural extension applicable to hybrid settings. To enable practical computation in many-body ground states, we developed a scalable numerical sampling algorithm for estimating both the Gaussian and hybrid magic R\'enyi entropies. We applied this method to compute the hybrid magic entropy using both perturbative and numerical Monte Carlo methods in the Dicke model and the Jaynes-Cummings model, prototypical spin-boson systems. We believe the work suggests a number of interesting future directions and we conclude with a brief discussion of these. 

Recent work~\cite{hoshino2025stabilizer} has identified universal scaling behaviour of the stabiliser R\'enyi entropy with certain boundary conditions in conformal field theory. It would be interesting to explore whether similar universal features arise in the Gaussian and hybrid magic R\'enyi entropies, and to study their scaling behaviour in bosonic critical systems---particularly at non-Gaussian fixed points. Along these lines, it would also be valuable to investigate the holographic interpretation of various magic entropy measures in the context of AdS/bCFT~\cite{fujita2011aspects}. 

Spin-fermion systems present another promising direction for future work since the Gaussian entropy is naturally extended to fermionic systems. One intriguing application is to the multi-channel Kondo impurity models, where conduction electrons interact with a local quantum spin, leading to partial screening and the emergence of anyonic quasiparticles. Computing the stabiliser R\'enyi entropy, fermionic Gaussian R\'enyi entropy, and hybrid spin-fermion R\'enyi entropy in these models may provide insights into the topological properties and quantum correlations associated with emergent anyons. Another interesting application is to study the interplay of supersymmetric localisation and magic entropy in supersymmetric quantum systems of bosons and fermions; the exactness of the saddle point approximation in this context is suggestive of non-trivial stabilizer properties.

\section*{Acknowledgements}
The work is supported by National Science and Technology Council of Taiwan under Grants 
No. NSTC 113-2112-M- 007-019, 114-2918-I-007-015. Numerical simulations in this project made extensive use of the open-source Python framework QuTiP \cite{johansson2012qutip,johansson2013qutip,lambert2024qutip5quantumtoolbox}. SC thanks Heng-Yu Chen (NTU) and Marco Paini (Rigetti) for many interesting discussions. P.-Y.C acknowledges support from Yukawa Institute for Theoretical Physics, Kyoto University, RIKEN Center for Interdisciplinary Theoretical and Mathematical Sciences, and National Center for Theoretical Sciences, Physics Division. Computational resources were provided by Academia Sinica Grid Computing Centre, supported by Grant No. AS-CFII-112-103.

\appendix

\section{Dicke model perturbation theory}\label{appendix:Dicke}
In this appendix we consider the perturbative computation of the quantity
\begin{align} \label{puremono}
  H_{\alpha}(p_{\psi}) = \frac{1}{(2^{N}\pi)^{\alpha}}\sum_{a,b}\int \mathrm{d}^{2}z \,|\bra{\psi}D(z)\otimes \sigma_{a,b}\ket{\psi}|^{2\alpha}
\end{align}

At $\lambda = 0$, the unperturbed ground state of the system is $\ket{\text{GS}(\lambda = 0)} = \ket{0}\ket{\downarrow}_{z}$, and the energy is $E_{0} = -\omega_{z}N$. For $\lambda \rightarrow 0$, the first order perturbation of the ground state is given by exciting the $i$-th spin from $\ket{\downarrow_{i}}$ to $\ket{\uparrow_{i}}$, i.e. $\sum_{i}\ket{1}\ket{\downarrow \cdots \uparrow_{i} \cdots \downarrow}_{z}$, where $E^{(0)} = \omega_{c} + 2\omega_{z} - \omega_{z}N$. We find to first order,
\begin{equation}
\begin{split}
  \ket{\lambda} &= \ket{0}\ket{\downarrow}_{z} - \lambda \bigg(\frac{2}{\sqrt{N}(\omega_{c} + 2\omega_{z})}\sum_{i}X_{i}\ket{1}\ket{\downarrow}_{z} \bigg) \\
  &\quad + \mathcal{O}(\lambda^{2}).
  \end{split}
\end{equation}
We may further normalise the state according to
\begin{equation}
  \ket{\lambda} = \mathcal{N}^{-1/2} \bigg(\ket{0}\ket{\downarrow}_{z} - \sqrt{\frac{\alpha}{N}}\sum_{i}X_{i}\ket{1}\ket{\downarrow}_{z} \bigg),
\end{equation}
where $\alpha = (\frac{2\lambda}{\omega_{c} + 2\omega_{z}} )^{2}$ and we denote the normalization by $\mathcal{N} = 1 + \alpha$. Elements of the Pauli group are denoted $P \subset \{W_{1}\otimes \cdots \otimes W_{N}\,|\, W_{i} = \{I,X,Y,Z\}_{i} \}$, and the displacement operator is defined as $D(z) \otimes P$, the non-zero subgroup of $P$ will be divided into three types $P_{I}$, $P_{II}$, and $P_{III}$
\begin{align}
\begin{split}
  P_{I} &= \{W_{1}\otimes \cdots \otimes W_{i} \otimes  \cdots \otimes W_{N}\}, \\
  P_{II} &= \{W_{1}\otimes \cdots \otimes T_{i}\otimes \cdots \otimes W_{N} \}, \\
  P_{III} &= \{W_{1}\otimes \cdots \otimes T_{i} \otimes T_{j} \otimes \cdots \otimes W_{N} \},
  \end{split}
\end{align}
where $W_{i}\in \{I,Z\}, \, T_{i} \in \{X,Y\}$. The first-type string $P_{I}$ flips or measure the states in $Z$ basis. The second-type string flips $\ket{\downarrow}_{z}$ to $\ket{\downarrow \cdots \uparrow_{i} \cdots \downarrow}_{z}$, vice versa. The third-type string $P_{III}$ interchange $\ket{\downarrow \cdots \uparrow_{i} \cdots \downarrow}_{z}$ between $\ket{\downarrow \cdots \uparrow_{j} \cdots \downarrow}_{z}$.

\subsection{Hybrid entropy}
Let us first introduce some notation. We write the Weyl function $\chi_{nm} = \bra{n}D(z)\ket{m}$, and consider
\begin{align}
\begin{split}
    \chi_{00} &= e^{\frac{-1}{2}|z|^{2}}, \quad \chi_{11} = e^{\frac{-1}{2}|z|^{2}}L_{1}(|z|^{2}), \\
    \chi_{01} &= z e^{\frac{-1}{2}|z|^{2}}, \quad  \!\!\!\chi_{10} = -z^{\star} e^{\frac{-1}{2}|z|^{2}}.
    \end{split}
\end{align}

We start by the type I string combined with the displacement operator $D(z)$ of the CV systems. Suppose there are $k$ $Z$ operators and $N-k$ $I$ operators. We denote the first-type string as $P_{I}$ and write
\begin{equation} \label{typeI}
  \chi_{I} = \bra{\lambda}D(z)\otimes P_{I} \ket{\lambda} = \frac{(-1)^{k}}{\mathcal{N}}\bigg(\chi_{00} +  \frac{(N - 2k)\alpha}{N}\chi_{11} \bigg),
\end{equation}
the number of $k$-$Z$ operators and $(N - k)$-$I$ operators is  $\binom{N}{k}$.

The second-type string $P_{II}$ is interchanging the two states, and the expectation value for $T_{i} = X_{i}$ is 
\begin{subequations}  \label{typeII}
\begin{align}
  \chi_{II,X} = \bra{\lambda}D(z)\otimes P_{II,X} \ket{\lambda} = \frac{(-1)^{k+1}}{\mathcal{N}} \sqrt{\frac{\alpha}{N}} ( \chi_{01} + \chi_{10} ),
\end{align}
and for $T_{i} = Y_{i}$, the string is 
\begin{align}
  \chi_{II,Y} = \bra{\lambda}D(z)\otimes P_{II,Y} \ket{\lambda} = i\frac{(-1)^{k+1}}{\mathcal{N}} \sqrt{\frac{\alpha}{N}}( \chi_{01} - \chi_{10}),
\end{align}
the two string has the same counting $\binom{N}{1}\binom{N-1}{k}$ for $k$-Z gates and $(N-1 - k)$-I gates.
\end{subequations}

The third-type string is more non-trivial than the previous two strings and is given by  
\begin{align} \label{typeIII}
 \chi_{III,XX} = \bra{\lambda}D(z)\otimes P_{III,XX} \ket{\lambda} = \frac{2\alpha}{N}\frac{(-1)^{k}}{\mathcal{N}} \chi_{11},
\end{align}
which is the same expectation value for $T_{i},T_{j} = YY$, i.e. $\bra{\lambda}D(z)\otimes P_{III,YY} \ket{\lambda} = \bra{\lambda}D(z)\otimes P_{III,XX} \ket{\lambda}$. For $XY$ cases, the amplitudes are zero since there exist two excited spins that cancel out. In the $XX$ and $YY$ cases, the counting numbers are both $\binom{N}{2}\binom{N-2}{k}$.

We now check that the summation and integration over all Pauli groups and bosonic degree of freedoms gives unity. The first-type string has the integration 
\begin{align}
\begin{split}
  \sum_{\{P_{I} \}}\int \frac{\mathrm{d}^{2}z}{2^{N}\pi} \,|\chi_{I}|^{2} &= \frac{1}{2^{N}\pi} \sum_{k} \binom{N}{k} \int \mathrm{d}^{2}z \, \frac{1}{\mathcal{N}^{2}}\bigg(|\chi_{00}|^{2}\\  & \quad + \frac{(N - 2k)^{2}\alpha^{2}}{N}|\chi_{11}|^{2}\bigg) \\
  &= \frac{N + \alpha^{2}}{N\mathcal{N}^{2}}.
  \end{split}
\end{align}
Here we drop the cross term in the following article since $\sum_{k}\binom{N}{k}(N - 2k)^{n} = 0$ when $n$ is odd.

The summation over all second-type string is more trivial, since $|\chi_{II,X}|^{2} = |\chi_{II,Y}|^{2}$ with $|\chi_{II}|^{2} = |\chi_{II,X}|^{2} + |\chi_{II,Y}|^{2}$ by the same odd-power argument above, 
\begin{align}
\begin{split}
 \sum_{\{P_{II} \}}&\int \frac{\mathrm{d}^{2}z}{2^{N}\pi}  \,|\chi_{II}|^{2} = 2\times \frac{1}{2^{N}\pi} \sum_{k} \binom{N}{1}\binom{N-1}{k} \\ &\quad \times \int \mathrm{d}^{2}z \, \frac{\alpha}{N\mathcal{N}^{2}}\bigg( 2|z|^{2} - z^{2} - (z^{\star})^{2} \bigg)e^{-|z|^{2}} \\
  &= \frac{2\alpha}{\mathcal{N}^{2}}.
 \end{split}
\end{align}
The last-type string has $|\chi_{III,XX}|^{2} = |\chi_{III,YY}|^{2} $ with $|\chi_{III}|^{2} = |\chi_{III,XX}|^{2} + |\chi_{III,YY}|^{2} $, where the summation is 
\begin{align}
\begin{split}
  \sum_{\{P_{III} \}}\int \frac{\mathrm{d}^{2}z}{2^{N}\pi}  \,|\chi_{III}|^{2} &= 2\times \frac{1}{2^{N}\pi}\sum_{k} \binom{N}{2}\binom{N-2}{k} \\ & \quad \times\int \mathrm{d}^{2}z \, \frac{4\alpha^{2}}{N^{2}\mathcal{N}^{2}}|\chi_{11}|^{2} \\
  &= \frac{N-1}{N}\frac{\alpha^{2}}{\mathcal{N}^{2}}.
  \end{split}
\end{align}
The summation over all types of Pauli strings gives unity, thus ensuring we obtain a valid probability density function.

Next we calculate the mutual magic of this perturbed ground state $\ket{\lambda}$, and seek to compute $H_{2}(p_{\lambda})$. For the first-type string, 
\begin{equation}
  \begin{split}
  |\chi_{I}|^{4} &= \mathcal{N}^{-4}\bigg[ |\chi_{00}|^{2} + \bigg(\frac{(N - 2k)\,\alpha}{N}\bigg)^{2}|\chi_{11}|^{2} \\ & \quad + \frac{2(N-2k)\,\alpha}{N}\chi_{00}\chi_{11} \bigg]^{2} \\
  &= \mathcal{N}^{-4}\bigg[ |\chi_{00}|^{4} + \bigg(\frac{(N - 2k)\,\alpha}{N}\bigg)^{4}|\chi_{11}|^{4} \\ &\quad + 6\bigg(\frac{(N-2k)\,\alpha}{N} \bigg)^{2}|\chi_{00}|^{2}|\chi_{11}|^{2} \bigg],
  \end{split}
\end{equation}
here we drop the term $(N - 2k)^{n}$ with $n$ odd, since by changing $(N - 2k) \rightarrow -(N - 2k)$ under $k \rightarrow N - k$, the terms involving odd powers go to zero. The summation then gives
\begin{align}
\begin{split}
  \sum_{k}\binom{N}{k}|\chi_{I}|^{4} &= 2^{N}\mathcal{N}^{-4}\bigg[ |\chi_{00}|^{4} + \frac{(N^{2} - 6N + 6)\alpha^{4}}{N^{3}}|\chi_{11}|^{4} \\ & \quad + \frac{6\alpha^{2}}{N} |\chi_{00}|^{2}|\chi_{11}|^{2} \bigg],
  \end{split}
\end{align}
we integrate out the CV variable to find
\begin{align} \label{puremonoI}
\begin{split}
  \int \mathrm{d}^{2}z \,\sum_{k}\binom{N}{k}|\chi_{I}|^{4} &= 2^{N}\mathcal{N}^{-4}\bigg[ \frac{\pi}{2} + \frac{(N^{2} - 6N + 6)\,\alpha^{4}}{N^{3}}\frac{\pi}{4} \\ & \quad + \frac{6\alpha^{2}}{N} \frac{\pi}{4} \bigg].
  \end{split}
\end{align}
For the second-type string, we use the polar coordinate $z = re^{i\theta}$ to write
\begin{align}
  |\chi_{II,X(Y)}|^{4} = \frac{\mathcal{N}^{-4}\alpha^{2}}{N^{2}}[2r^{2} \pm 2r^{2}\cos(2\theta)]^{2}e^{-2r^{2}},
\end{align}
and the summation over Pauli strings ($X$ or $Y$) gives  a factor $N2^{N-1}$, and then we integrate over phase space where the second-type string contribution for $H_{2}(p_{\lambda})$ is 
\begin{align} \label{puremonoII}
  \int r\,\mathrm{d}r \,\mathrm{d}\theta \, N2^{N-1}(|\chi_{II,X}|^{4} + |\chi_{II,Y}|^{4}) = \frac{2^{N-1}}{\mathcal{N}^{4}} \bigg( \frac{3\pi \alpha^{2}}{N} \bigg).
\end{align}
For the last type string, the generalized Weyl functions for either XX and YY-type strings are given by 
\begin{align}
  |\chi_{III,XX}|^{4} = |\chi_{III,YY}|^{4} = \frac{16\alpha^{4}\mathcal{N}^{-4}}{N^{4}}|\chi_{11}|^{4},
\end{align}
and the the third-type string contribution for $H_{2}(p_{\lambda})$ is 
\begin{align} \label{puremonoIII}
  \int \mathrm{d}^{2}z \, \frac{N(N-1)2^{N-2}}{2} \, \bigg( \frac{32\alpha^{4}}{N^{4}\mathcal{N}^{4}}|\chi_{11}|^{4} \bigg) = \frac{2^{N}}{\mathcal{N}^{4}}\bigg(\frac{(N-1)\alpha^{4}\pi}{N^{3}}\bigg).
\end{align}
Finally, summing the contributions \eqref{puremonoI}, \eqref{puremonoII}, and \eqref{puremonoIII}, we find the pure state entropy for the perturbed ground state in this case is 
\begin{align}
  \begin{split}
  H_{2}(p_{\psi}) &= -\log\bigg[\frac{1}{2^{N}}\frac{1}{2\pi}\mathcal{N}^{-4}\bigg( 1 + \frac{1}{2}\frac{(N^{2} - 4N + 4)\alpha^{4}}{N^{3}} + \frac{6\alpha^{2}}{N} \bigg)\bigg] \\
  &= N\log(2) + \log(2\pi) + 4\log(\mathcal{N}) \\ & \quad - \log\bigg( 1 +  \frac{6\alpha^{2}}{N} + \frac{1}{2}\frac{(N^{2} - 2N + 2)\alpha^{4}}{N^{3}} \bigg).
  \end{split}
\end{align}

\subsection{Gaussian entropy--tracing over spin}
The density matrix for the system is 
\begin{align} \label{mixDM}
\begin{split}
  \rho = \ketbra{\lambda}{\lambda} &= \mathcal{N}^{-1}\bigg(\ket{0}\ket{\downarrow}-\sqrt{\frac{\alpha}{N}}\ket{1}\sum_{i}X_{i}\ket{\downarrow}\bigg) \\ & \quad \otimes \bigg(\bra{0}\bra{\downarrow}-\sqrt{\frac{\alpha}{N}}\bra{1}\sum_{i}X_{i}\bra{\downarrow}\bigg),
  \end{split}
\end{align}
tracing out the spin degrees of freedom gives 
\begin{align}
  \rho_{b} = \Tr_{s}(\rho) = \mathcal{N}^{-1}(\ketbra{0}{0} + \alpha \ketbra{1}{1}),
\end{align}
the reduced density matrix for the Bosonic mode is not a pure state, as the purity is $\Tr(\rho_{b}^{2}) = \mathcal{N}^{-2}(1 + \alpha^{2}) < 1$. To compute the Weyl probability distribution we calculate the expectation value of the displacement operator $D(z)$
\begin{align}
 \langle D(z) \rangle_{\rho_{b}} =  \Tr(\rho_{b} D(z)) = \mathcal{N}^{-1}(\chi_{00} + \alpha \chi_{11}).
\end{align}
The \textit{mixed state GRE} is then given by (for $\alpha = 2$)
\begin{align}
  \begin{split}
  G_{2}(\rho_{b}) &= -\log(\int \mathrm{d}^{2}z \, p^{2}(\rho_{b})) \\
  &= \log(2\pi) + 2\log( \Tr(\rho_{b}^{2})) + 4\log(\mathcal{N}) \\
  & \quad - \log(1 + 2\alpha + 3\alpha^{2} + \alpha^{3} + \frac{1}{2}\alpha^{4}).
  \end{split}
\end{align}

\subsection{Stabilizer entropy--tracing over boson}
The reduced density matrix for the spin degrees of freedom to first order is
\begin{align}
  \rho_{s} = \Tr_{b}(\rho) = \mathcal{N}^{-1}\bigg( \ketbra{\downarrow}{\downarrow} + \frac{\alpha}{N}\sum_{i,j}X_{i}\ketbra{\downarrow}{\downarrow}X_{j} \bigg),
\end{align}
similar to the pure state calculation, we focus on the expectation value of each Pauli-strings, and examine whether the sum of all the Pauli-strings configurations gives unity. First, similar to the pure state monotone, we start by the first-type string
\begin{align}
  \langle P_{I} \rangle_{\rho_{s}} = \mathcal{N}^{-1} \bigg( 1 + \frac{\alpha (N-2k)}{N}  \bigg),
\end{align}
for the second-type string, the expectation values of Pauli-strings are all zero, as they can be viewed as both projection unto other states, and so provide zero diagonal matrix element. However, for the last type of string, as the Pauli string may match the projection $\ketbra{\downarrow \cdots \uparrow_{i} \cdots \downarrow }{\downarrow \cdots \uparrow_{j} \cdots \downarrow}$ to $\ketbra{\downarrow \cdots \uparrow_{i} \cdots \downarrow }{\downarrow \cdots \uparrow_{i} \cdots \downarrow}$ and $\ketbra{\downarrow \cdots \uparrow_{j} \cdots \downarrow }{\downarrow \cdots \uparrow_{j} \cdots \downarrow}$ (for $i \neq j$), and the expectation value is non-zero
\begin{align}
  \langle P_{III} \rangle_{\rho_{s}} = \mathcal{N}^{-1} \bigg( \frac{2\alpha}{N} \bigg),
\end{align}
the factor $2$ comes from the fact that one Pauli string can have two non-zero diagonal matrix elements.

In this case, the sum of all Pauli strings is 
\begin{align}
  \frac{1}{2^{N}}\sum_{P \in \mathcal{P}}\langle P \rangle_{\rho_{s}}^{2} = \Tr(\rho_{s}^{2}) = \Tr(\rho_{b}^{2}) .
\end{align} 
Now we compute, the stabilizer R\'enyi entropy with $\alpha = 2$
\begin{align}
  \begin{split}
  M_{2}(\rho_{s}) &= -\log\bigg[ \frac{\mathcal{N}^{-4}}{4^{N}\Tr(\rho_{s}^{2})} \bigg( \sum_{k}\binom{N}{k}\langle P_{I} \rangle_{\rho_{s}}^{4} \\
  &\quad + 2\times \sum_{k}\binom{N-2}{k}\binom{N}{2}\bigg(\frac{2\alpha}{N}\bigg)^{4} \bigg)\bigg] \\
  &= N\log(2) + 4\log(\mathcal{N}) + 2\log( \Tr(\rho_{s}^{2})) \\
  & \quad - \log\bigg( 1 + \frac{6\alpha^{2}}{N} + \frac{(N^{2} - 2N + 2)\alpha^{4}}{N^{3}} \bigg).
  \end{split}
\end{align}

\bibliography{magicentropy}

\end{document}